\documentclass[10pt,journal,compsoc]{IEEEtran}

\usepackage{comment}
\usepackage{amssymb} 
\usepackage{color}
\usepackage{booktabs}

\usepackage{multirow}
\usepackage{wrapfig}
\usepackage{makecell}
\usepackage{bbding}
\usepackage{adjustbox}
\usepackage{array}
\usepackage{xcolor,colortbl}
\usepackage{balance}

\usepackage{hyperref}
\usepackage[hyphenbreaks]{breakurl}

\newcommand{\Bmat}{{\bf B}}

\newcommand{\Dmat}{{\bf D}}

\newcommand{\Hmat}[0]{{{\bf H}}}

\newcommand{\Kmat}[0]{{{\bf K}}}

\newcommand{\Mmat}[0]{{{\bf M}}}

\newcommand{\Qmat}[0]{{{\bf Q}}}
\newcommand{\Rmat}[0]{{{\bf R}}}

\newcommand{\Vmat}[0]{{{\bf V}}}
\newcommand{\Wmat}[0]{{{\bf W}}}
\newcommand{\Xmat}{{\bf X}}
\newcommand{\Ymat}[0]{{{\bf Y}}}
\newcommand{\Zmat}{{\bf Z}}

\newcommand{\xv}{\boldsymbol{x}}
\newcommand{\yv}{\boldsymbol{y}}

\newcommand{\zv}{\boldsymbol{z}}

\newcommand{\ie}{{\em i.e.}}
\newcommand{\eg}{{\em e.g.}}

\ifCLASSOPTIONcompsoc
  \usepackage[nocompress]{cite}
\else
  \usepackage{cite}
\fi
\usepackage{amsmath}
\begin{document}
%
\title{Spatial-Temporal Transformer for Video Snapshot Compressive Imaging}
%
%
%
%

\author{Lishun~Wang, 
Miao~Cao,
        Yong Zhong
        and~Xin~Yuan,~\IEEEmembership{Senior~Member,~IEEE}
\IEEEcompsocitemizethanks{\IEEEcompsocthanksitem L. Wang and Y. Zhong are with Chengdu Institute of Computer Application Chinese Academy of Sciences, Chengdu, Sichuan 610041, China and also with University of Chinese Academy of Sciences, Beijing 100049, China.\protect\\
E-mails: wanglishun17@mails.ucas.edu.cn, zhongyong@casit.com.cn.\protect
\IEEEcompsocthanksitem M. Cao and X. Yuan are with Westlake University, Hangzhou 310024, China.  E-mail: \{caomiao, xyuan\}@westlake.edu.cn. \protect
\IEEEcompsocthanksitem Corresponding author: Xin Yuan.
}
\thanks{Manuscript updated \today.}
}

\IEEEtitleabstractindextext{%
\begin{abstract}
Video snapshot compressive imaging (SCI)  captures multiple sequential video frames by a single measurement using the idea of computational imaging. The underlying principle is to modulate high-speed frames through different masks and these modulated frames are summed to a single measurement captured by a low-speed 2D sensor (dubbed optical encoder); following this, algorithms are employed to reconstruct the desired high-speed frames (dubbed software decoder) if needed.
In this paper, we consider the reconstruction algorithm in video SCI, \ie, recovering a series of video frames from a compressed measurement. Specifically, we propose a Spatial-Temporal transFormer (STFormer) to exploit the correlation in both spatial and temporal domains. STFormer network is composed of a token generation block, a video reconstruction block, and these two blocks are connected by a series of STFormer blocks. 
Each STFormer block consists of a spatial self-attention branch, a temporal self-attention branch and the outputs of these two branches are integrated by a fusion network.
Extensive results on both simulated and real data demonstrate the state-of-the-art performance of STFormer. The code and models are publicly available at ~\url{https://github.com/ucaswangls/STFormer}
\end{abstract}
}

\maketitle

\IEEEdisplaynontitleabstractindextext

%
\IEEEpeerreviewmaketitle

\ifCLASSOPTIONcompsoc
\IEEEraisesectionheading{\section{Introduction}\label{sec:introduction}}
\else
\section{Introduction\label{Sec:Intro}}

\fi
\IEEEPARstart{W}{ith} recent advances in artificial intelligence, 
{high quality, high-dimensional data have become one of the bottlenecks for large-scale deep learning models.} 
In other words, capturing more data with multiple dimensions will lead to a dramatic increase in storage and transmission costs. 
Unlike ordinary cameras which capture RGB images, 
computational imaging~\cite{Mait18CI,Altmann18Science} provides a new way to capture high-dimensional data in a memory-efficient manner.
{It is  promising to support the explosion of artificial intelligence using data captured by computational imaging systems.}
In this paper, we focus on snapshot compressive imaging (SCI)~\cite{Yuan2021_SPM}, especially video SCI systems, 
which are capable of capturing high-speed videos using a low-speed camera~\cite{Llull2013}, enjoying the benefits of low memory requirement, low bandwidth for transmission, low cost and low power.
\begin{figure}[!ht]
  \centering 
  \includegraphics[width=1.\linewidth]{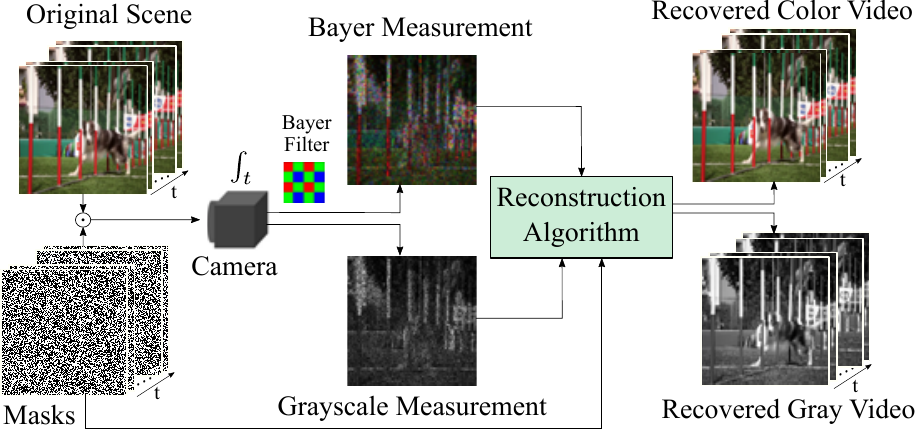}
  \caption{Schematic diagram of grayscale and color video SCI. 
  A series of original frames are modulated through different masks, 
  and then a camera is used to integrate encoded frames 
  to obtain the compressed measurement, which is convenient for storage and transmission. 
  Then, the measurement and masks are input into the reconstruction algorithm to recover 
  the original video frames~\cite{Yuan2014}.}
  \label{fig:sci}
\end{figure}

\subsection{Video Snapshot Compressive Imaging}
Traditional high-speed camera method for capturing high-speed scenes 
often faces the disadvantages of high hardware cost, high storage requirement and high transmission bandwidth. 
Bearing these challenges, video SCI system provides an elegant solution. 
As shown in Fig.~\ref{fig:sci}, video SCI is an integrated hardware plus software system. 
For the hardware part (encoder), each frame of the original video is encoded by different masks, 
then a series of encoded frames is integrated (summed) by the (grayscale or color) camera to generate a compressed measurement. 
In this manner, video SCI can achieve efficient compression during optical domain imaging and 
improve the efficiency of video storage and transmission.
At present, a variety of video SCI systems \cite{Llull2013,Hitomi2011,Reddy2011,Qiao2020,Sun2017} have been built.  
The mask is usually generated by a digital micromirror device (DMD) or a spatial light modulator (SLM), 
and the encoded measurement is usually captured by a charge-coupled device (CCD) or a
complementary metal-oxide semiconductor (CMOS) camera. 
In addition, for color video SCI \cite{Yuan2014}, there is usually a Bayer-filter before the sensor array to capture different color components such as red, green or blue at different pixels, 
and ultimately the camera outputs the modulated Bayer measurement. 
For the software part (decoder), measurement and masks are usually fed into the reconstruction algorithm to recover the desired high-speed video.

\subsection{Reconstruction Algorithms for Video SCI}
In the decoding stage, the video SCI reconstruction algorithm aims to solve an ill-posed inverse problem. 
Traditional model-based reconstruction algorithms combines 
the idea of iterative optimization, such as generalized alternating projection (GAP) \cite{Liao2014}
and alternating direction method of multipliers (ADMM) \cite{Boyd2011}, 
with a variety of prior knowledge, 
such as total variation (TV) \cite{Yuan2016}, non-local low rank \cite{Liu2018} and Gaussian mixture model \cite{Yang2014}.
Although these methods do not require any data to train the model, 
they usually suffer from poor reconstruction quality and long reconstruction time. 
In recent years, due to the development of deep learning and its strong generalization ability, 
researchers have constructed many learning-based models. 
For example, BIRNAT \cite{Cheng2020b} uses convolutional neural networks (CNNs) and bidirectional recurrent neural networks (RNNs) for frame-by-frame reconstruction, 
while U-net \cite{Qiao2020} uses a simple U-shaped structure for fast reconstruction. 
RevSCI \cite{Cheng2021} can save training memory by using reversible mechanism \cite{behrmann2019invertible}. Plug-and-play (PnP) methods, such as PnP-FFDNet \cite{Yuan2020c} and PnP-FastDVDnet \cite{yuan2021plug}, 
make the model more flexible by integrating 
the deep denoising model into an iterative optimization algorithm. 
GAP-net \cite{meng2020gap}, DUN-3DUnet \cite{Wu2021} and Two-stage \cite{zheng2022two} 
further improve reconstruction quality and speed by using the ideas of deep unfolding~\cite{Yang2022}. 
Although previous methods have made great strides in video SCI reconstruction, 
at present, video SCI reconstruction still suffers from several key challenges:  
\begin{itemize}
    \item[1)] 
   {
  Since the video SCI sampling process compresses the entire temporal domain, it is important to establish long-term temporal correlation in the temporal domain during reconstruction. However, 
  existing networks \cite{saideni2022overview} use convolution to explore temporal correlation, and its receptive field are too small to be suitable for long-term temporal correlation extraction, resulting in poor reconstruction quality in complex high-speed scenes.}
\item [2)] Most existing end-to-end deep learning methods can not adapt to masks and input sizes changes. 
More specifically, the model usually needs to be retrained or fine-tuned when masks and input sizes change~\cite{Wang2021e}.
\item [3)] PnP methods usually suffer from excessive smoothing and loss of local details.
\end{itemize}

{
Bearing these concerns in mind, we leverage the powerful model
capacity of Transformer \cite{transformer} and its ability to explore long-term dependencies to build an efficient video SCI reconstruction network. 
}
\subsection{Contributions of This Paper}
In this paper, we propose a simple yet effective network for video SCI reconstruction. Our main contributions are as follows:

\begin{itemize} 
  \item We build an end-to-end deep video SCI reconstruction network 
  based on spatial-temporal Transformer, 
  which can be well applied to multiple video SCI reconstruction tasks.

 \item Through {\em space-time factorization} and local self-attention  mechanism, 
  we propose an {\em efficient and flexible Transformer} module, dubbed
   STFormer, with {\bf linear complexity}, 
  which can explore spatial-temporal correlations efficiently.

\item We propose a {\bf Grouping Resnet Feed Forward} module to 
  further improve reconstruction quality by fusing multi-layer information 
  and strengthening the correlation between different layers.
\item We use multiple 3D convolutions to construct a token generation block to prevent loss of local details. 
\item Experimental results on a large number of simulated  
  and real datasets demonstrate that our proposed algorithm achieves state-of-the-art  (SOTA) results 
  with better real-time performance compared to previous methods.
\end{itemize} 

The rest of this paper is organized as follows: 
Sec.~\ref{Sec:Related} introduces the mathematical model of grayscale and color video SCI and the related work of
vision Transformers. 
In Sec.~\ref{Sec:ProSTF}, we propose a reconstruction model STFormer for different video SCI reconstruction tasks, including grayscale video, color video and large-scale video.
Sec.~\ref{Sec:Result} shows experimental results on multiple simulated and real datasets. 
Sec.~\ref{Sec:Con} concludes the entire paper.

\section{Related Work \label{Sec:Related}}
Since video SCI is an integrated system of hardware and software, in the following, 
we first introduce the mathematical model of video SCI ,
and then briefly review the related work of vision Transformer, 
which is the main part of the reconstruction network. 

\subsection{Review of Mathematical Model for video SCI}
Video SCI system encodes high-dimensional video data into a 2D measurement, and 
coded aperture compressive temporal imaging (CACTI) \cite{Llull2013} is one of the earliest video SCI systems. 
As shown in Fig.~\ref{fig:sci}, the three-dimensional video data is first modulated by multiple masks. 
Then, the encoded high-speed scene is captured by a two-dimensional camera sensor 
through integration across the time dimension.

For grayscale video SCI system, let $\Xmat\in\mathbb{R}^{n_x\times{n_y}\times{B}}$ denote the $B$-frame (grayscale) video data to be captured, 
$\Mmat\in\mathbb{R}^{n_x\times{n_y}\times{B}}$ denote pre-defined masks.
For each frame within the video $f=1,\ldots,B$, it is modulated by a mask in the image plane conducted by a 4-f system, and we can express this modulation as
\begin{equation}
  \Xmat^{'}(:,:,f) = \Xmat(:,:,f)\odot\Mmat(:,:,f),  
\end{equation}
where $\Xmat^{'}\in\mathbb{R}^{n_x\times{n_y}\times{B}}$
denotes the modulated video data,
$\Xmat(:,:,f)$ denotes the $f$-th frame of the 3D video data $\Xmat$, 
$\Mmat(:,:,f)$ denotes the $f$-th mask of the 3D mask $\Mmat$,
and $\odot$ denotes the element-wise multiplication. 

After this, by compressing across the time domain,  
the camera sensor plane captures a 2D compressed measurement 
$\Ymat\in\mathbb{R}^{n_x\times{n_y}}$, which can be expressed as
\begin{equation}
  \Ymat=\sum_{f = 1}^{B}{\Xmat^{'}(:,:,f)+\Zmat},
  \label{eq:Y_mat}
\end{equation}
where $\Zmat\in\mathbb{R}^{n_x\times{n_y}}$ 
denotes the measurement noise. 

To keep the notations simple, we further give the vectorized formulation expression of Eq. \eqref{eq:Y_mat}. 
Firstly, we define $\operatorname{vec}(\cdot)$ as a vectorization operation of ensued matrix.
Then we vectorize 
\begin{eqnarray}
\yv&=&\operatorname{vec}(\Ymat)\in\mathbb{R}^{n_x{n_y}},\\
\zv&=&\operatorname{vec}(\Zmat)\in\mathbb{R}^{n_x{n_y}},\\
\xv&=&\left[\xv_1^{\top},\ldots,\xv^{\top}_{B}\right]^{\top}\in\mathbb{R}^{n_x{n_y{B}}},
\end{eqnarray}
where $\xv_f=\operatorname{vec}(\Xmat(:,:,f))$. 
In addition, we define sensing matrix generated by masks in SCI system as
\begin{equation}
  \Hmat = \left[\Dmat_1,\ldots,\Dmat_{B}\right]\in\mathbb{R}^{n_x{n_y}\times{n_x{n_y{B}}}}, 
  \label{eq:H}
\end{equation}
where $\Dmat_f=\operatorname{Diag}(\operatorname{vec}(\Mmat_f)) \in \mathbb{R}^{n_{x} n_{y} \times n_{x} n_{y}}$ is a diagonal matrix and its diagonal elements is filled by $\operatorname{vec}(\Mmat_f)$, 
Finally, the vectorization expression of Eq. (\ref{eq:Y_mat}) is
\begin{equation}
  \yv = \Hmat{\xv}+\zv. 
  \label{eq:y_vec}
\end{equation}

After obtaining the measurement $\yv$ (captured by the camera), the next task is 
to develop a decoding algorithm, \ie, given $\yv$ and $\Hmat$, solve $\xv$.

For color video SCI system, we use the Bayer pattern filter sensor, 
where each pixel captures only the red (R), green (G) or blue (B) channel of the raw data in a spatial layout such as `RGGB'. Since the colors of adjacent pixels are discontinuous, 
we divide the original measurement $\Ymat$ into four sub-measurements 
$\{\Ymat^{r},\Ymat^{g1},\Ymat^{g2},\Ymat^{b}\}\in\mathbb{R}^{\frac{n_x}{2}\times{\frac{n_y}{2}}}$. 
Correspondingly, the original masks $\Mmat$ and original video frames $\Xmat$ can also be divided into 
$\{\Mmat^{r},\Mmat^{g1},\Mmat^{g2},\Mmat^{b}\}\in\mathbb{R}^{\frac{n_x}{2}\times{\frac{n_y}{2}}\times{B}}$ and 
$\{\Xmat^{r},\Xmat^{g1},\Xmat^{g2},\Xmat^{b}\}\in\mathbb{R}^{\frac{n_x}{2}\times{\frac{n_y}{2}}\times{B}}$, respectively. 
The forward model of each channel can be expressed as
\begin{eqnarray}
  \Ymat^{r} &=& \sum_{f = 1}^{B}{\Xmat^{r}_f\odot\Mmat^{r}_f+\Zmat^{r}}, \\
  \Ymat^{g1} &=& \sum_{f = 1}^{B}{\Xmat^{g1}_f\odot\Mmat^{g1}_f+\Zmat^{g1}}, \\
  \Ymat^{g2} &=& \sum_{f = 1}^{B}{\Xmat^{g2}_f\odot\Mmat^{g2}_f+\Zmat^{g2}}, \\
  \Ymat^{b} &=& \sum_{f = 1}^{B}{\Xmat^{b}_f\odot\Mmat^{b}_f+\Zmat^{b}}.
\end{eqnarray}
For reconstruction, most previous algorithms \cite{Yuan2014,Yuan2020c} reconstruct each sub-measurement separately, 
and then use off-the-shelf demosaic algorithms to obtain the desired RGB color videos.
This channel-separated reconstruction algorithm cannot make good use of the correlation between channels and is inefficient.  
Therefore, we input all sub-measurements into our proposed reconstruction network simultaneously, 
and the reconstruction network directly outputs the desired RGB color video 
without employing previous demosaic algorithms.

\subsection{Vision Transformers \label{Sec:ViT}}  
Compared with convolutional networks \cite{Krizhevsky2012,He2016,Huang2017}, Vision Transformer (ViT) \cite{Dosovitskiy2020} 
and its variants \cite{Dong2021a,Zhu2020,Yuan2021b,wang2021pyramid,han2022survey}
have achieved competitive results in the field of image classification \cite{Deng2009}.
However, the original ViT model training requires a large dataset 
(\eg, JFT-300M~\cite{sun2017revisiting}) to achieve good results, 
DeiT \cite{touvron2021training} introduced some training strategies to enable the ViT model to achieve better performance in the ImageNet-1K dataset \cite{Deng2009}, 
but it still used the global self-attention mechanism, 
and the computational complexity increases quadratically with the image size. 
These greatly limit the application of ViT to dense prediction tasks 
such as object detection \cite{Lin2014}, semantic segmentation \cite{Zhou2017,Zhou2019,huang2019ccnet},
and low-level tasks such as image denoising \cite{zhang2017} and image restoration \cite{Liang2021a}. 

Moreover, PVT V2 \cite{wang2022pvt} proposed a linear spatial reduction attention (SRA) layer, 
which uses average pooling to reduce spatial dimension, and achieve improvements on fundamental vision tasks such as classification, detection, and segmentation. 
MST \cite{cai2022mask} proposed a spectral-wise multi-head self-attention (S-MSA) and customized a mask-guided mechanism (MM), which effectively explores the application of Transformer to the spectral SCI task \cite{meng2021self}. 
Swin Transformer \cite {Liu2021} limits the self-attention calculation to the local window 
through the division of non-overlapping local windows and shifted window mechanism, 
which reduces the computational complexity of the Transformer to a linear relationship with the image size. 
However, Swin Transformer can only explore spatial correlation. 
Video Swin Transformer \cite{ze2021video} can obtain local-scale temporal correlation by extending the local window to the temporal domain, 
but still cannot achieve good results in long-term correlation scenarios. 
TimeSformer \cite{bertasius2021space} investigates the spatial-temporal correlation between tokens through a space-time factorization mechanism, 
and has achieved extremely competitive results in the field of video recognition. 
However, due to the use of global self-attention mechanism in the spatial domain, 
the computational complexity grows quadratically with the image size, 
and its positional embedding size is fixed, lacking the flexibility to change with input size. 
In addition, most of the existing Transformers \cite{wang2022bevt,he2022masked,bao2021beit} divide images or videos into non-overlapping patches to generate tokens, 
resulting in loss of local details and cannot be well applied to video reconstruction tasks in this work.

\begin{figure*}[!ht]
  \centering 
  \includegraphics[width=1.\linewidth]{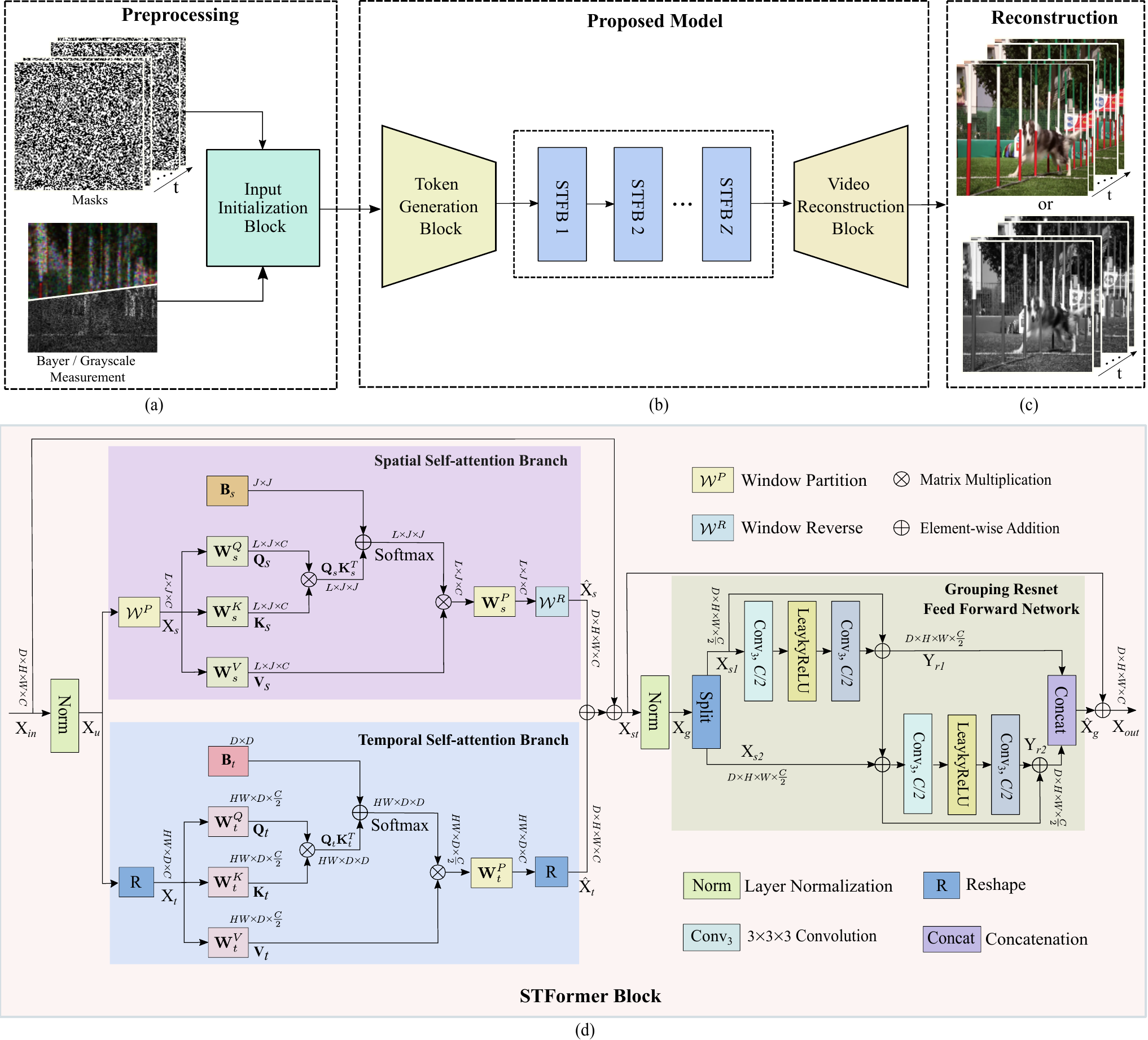}
  \caption{Architecture of the proposed STFormer and the overall process of color or grayscale video reconstruction. 
  (a) After obtaining measurement and masks, the preprocessing stage generates the estimated value 
  of the modulated frames through an input initialization block, 
  and uses the estimated value as the input to the STFormer network for reconstruction.
  (b) The proposed STFormer network is composed of a token generation block, $Z$ STFormer blocks, and a video reconstruction block.
  (c) Reconstructed color or grayscale video. 
  (d) Details of STFormer block, mainly composed of spatial self-attention branch, temporal self-attention branch and grouping resnet feed forward network, where $J=G_h{G_w}$ represents the number of tokens in each local window, 
  $L=D\frac{HW}{J}$  represents the number of local windows, and $G_h$, $G_w$ represent the height and width of the local window, respectively. For the convenience of presentation, only the heads $N=1$ scenario is described here. Please refer to Sec.~\ref{sec:stformer} for details.}
  \label{fig:network}
\end{figure*}

\begin{figure}[!htbp]
  \centering 
  \includegraphics[width=.8\linewidth]{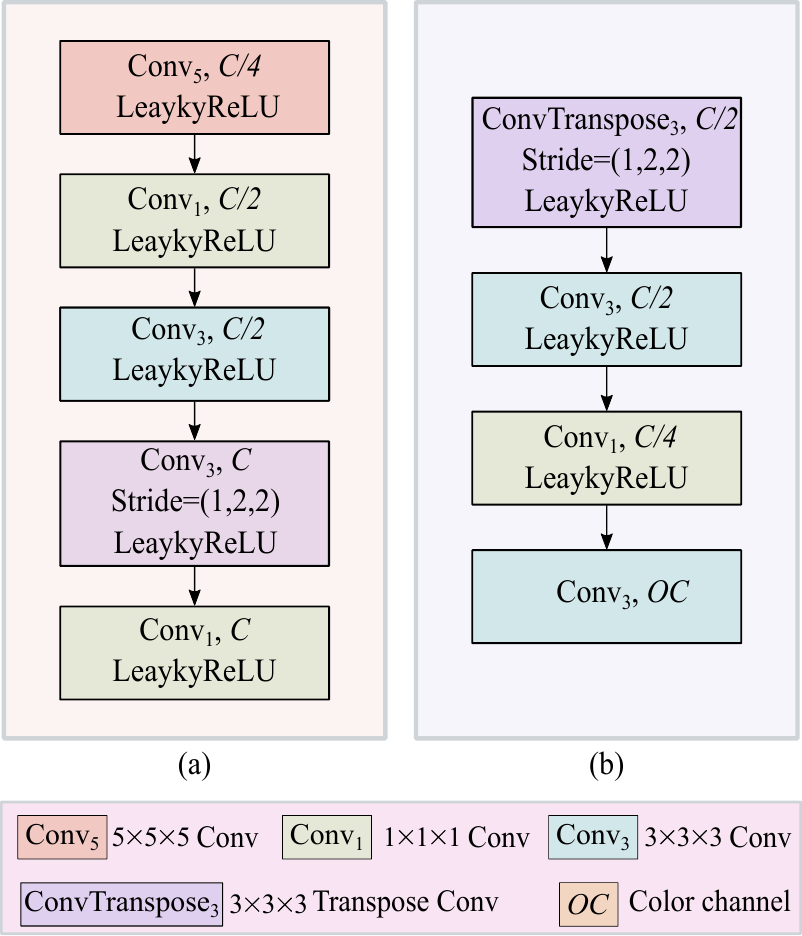}
  \caption{(a) Token generation block. (b) Video reconstruction block. }
  \label{fig:token}
\end{figure}

\section{Proposed STFormer Network for Video SCI \label{Sec:ProSTF}}
Inspired by recent advances in Transformers, especially variants for video processing tasks, 
in this section, we introduce our proposed spatial-temporal Transformer in detail, 
which can efficiently explore long-range spatial-temporal correlations and lead to SOTA results on video SCI tasks.

\subsection{Overall Architecture}
The overall video reconstruction network architecture is depicted in Fig.~\ref{fig:network}.  
In the preprocessing stage shown in Fig.~\ref{fig:network}(a), 
{
  we design the input initialization block with reference to GAP-net \cite{meng2020gap} and Two-stage \cite{zheng2022two}. 
  By pre-processing masks ($\Mmat$) and measurement ($\Ymat$) with the input initialization block, 
  we can get a coarse estimation of the modulated frames $\hat{\Xmat}\in\mathbb{R}^{n_x\times{n_y}\times{IC}\times{B}}$, where $IC=1$ represents the input channel number. 
  Then, $\hat{\Xmat}$ is fed into the token generation (TG) block \cite{Cheng2021} to get a series of consecutive tokens.
  We show in the supplementary material the implementation details of the pre-processing for grayscale and color video SCI in the input initialization block.  
}

As demonstrated in Fig.~\ref{fig:token}(a), the TG block consists of five 3D convolutional layers, 
each followed by a LeakyReLU activation function \cite{Xu2015}. 
Through the stride step and feature mapping in the convolutional layer, 
the number of finally generated tokens is $\frac{n_x}{2}\times\frac{n_y}{2}\times{B}$, and the feature dimension of each token is $C$. 
Different from most token generation methods, 
our proposed approach does not divide $\hat{\Xmat}$  into non-overlapping patches, 
but {\em uses 3D convolution for feature mapping}, 
and then {\em treats each point of the feature map as a token}. 
This is beneficial to reduce the phenomenon of local detail loss. 

To better explore the spatial-temporal correlation between each token, 
we design STFormer block (described in Section \ref{sec:stformer}), and stack $Z$ STFormer blocks. 
It is worth noting that we do not use any downsampling method in each STFormer block, 
and the input and output dimensions of each STFormer block are kept consistent, 
which is also beneficial to prevent the loss of local details. 

After the original tokens been mapped by $Z$ STFormer blocks, 
the spatial-temporal correlation between tokens has been well established. 
We input the tokens into the video reconstruction (VR) block \cite{Cheng2021} as shown in Fig.~\ref{fig:token}(b) 
to obtain the final desired video frames. 
In the VR block, the output channels number $OC$ of the last layer of convolution varies according to the reconstruction task, 
which is 1 for grayscale video reconstruction, 
and 3 for color video reconstruction.

\subsection{STFormer Block}
\label{sec:stformer}
As shown in Fig.~\ref{fig:network} (d), the STFormer block is composed of three parts: spatial self-attention (SSA) branch, 
temporal self-attenion (TSA) branch, and grouping resnet feed forward (GRFF) network. 
Among them, the SSA block can well explore the spatial correlation between tokens, 
and TSA block can well establish the temporal correlation between tokens, respectively. 
GRFF network can further investigate the correlation between adjacent tokens.

\subsubsection{Spatial Self-attention Branch}
Most of previous Transformer architectures used a global self-attention mechanism 
to calculate the correlation between tokens, 
which will cause the computational complexity of the Transformer model 
to increase quadratically with the number of tokens. This limits the application of Transformer to video SCI reconstruction tasks. 

Bearing this in mind, in our proposed spatial self-attention branch, 
we use a local self-attention method to calculate the spatial correlation.
As shown in Fig.~\ref{fig:network} (d), in SSA branch, we first divide the feature map $\Xmat_{u}\in\mathbb{R}^{D\times{H}\times{W}\times{C}}$ 
into a series of non-overlapping local windows $\Xmat_s\in\mathbb{R}^{L\times{J}\times{C}}$, 
where $J=G_h{G_w}{G_d}$ represents the number of tokens in each local window, 
$L=D\frac{HW}{J}$ represents the number of local windows, and  
$G_h, G_w, G_d$ represent the height, width and timing length of the local window, respectively. 
Since spatial self-attention only calculates the spatial correlation between tokens, 
we set $G_d$ to 1, and the default values of $G_h$ and $G_w$ are 7.
Then, the self-attention computation is restricted to each non-overlapping local window. 

For the self-attention calculation of the local window, we first linearly map $\Xmat_s$ to get 
$query\;\Qmat_s\in\mathbb{R}^{L\times{J}\times{C}}$,
$key\;\Kmat_s\in\mathbb{R}^{L\times{J}\times{C}}$ and $value\;\Vmat_s\in\mathbb{R}^{L\times{J}\times{C}}$: 
\begin{eqnarray}
  \Qmat_s &=& \Xmat_s\Wmat^Q_s, \\
  \Kmat_s &=& \Xmat_s\Wmat^K_s, \\
  \Vmat_s &=& \Xmat_s\Wmat^V_s,
\end{eqnarray}
where $\Wmat^Q_s\in\mathbb{R}^{C\times{C}},\Wmat^K_s\in\mathbb{R}^{C\times{C}}$ and $\Wmat^V_s\in\mathbb{R}^{C\times{C}}$ represent projection matrices and share parameters between different windows. 
Then, we respectively divide $\Qmat_s$, $\Kmat_s$, $\Vmat_s$ into $N$ heads along the feature channel $C$: 
\begin{eqnarray}
  \Qmat_s&=&[\Qmat_s^1,\cdots,\Qmat_s^N],\\
  \Kmat_s&=&[\Kmat_s^1,\cdots,\Kmat_s^N],\\
  \Vmat_s&=&[\Vmat_s^1,\cdots,\Vmat_s^N],
\end{eqnarray}
and the feature dimension of each head becomes $d_h=\frac{C}{N}$. 
For each head $i=1,\cdots,N$, the attention can be calculated by the local self-attention mechanism as: 
\begin{equation}
  {\rm Attention}(\Qmat_s^i,\Kmat_s^i,\Vmat_s^i)={\rm SoftMax}(\Qmat_s^{i}{\Kmat_s^{i}}^T\sqrt{d}+\Bmat_s^i)\Vmat_s^i,
\end{equation}
where ${\Kmat_s^{i}}^T$ represents the transpose of matrix $\Kmat_s^{i}$ 
and $\Bmat_s^i\in\mathbb{R}^{J\times{J}}$ represents the learnable relative position encoding. 
After this, we concatenate the outputs of $N$ heads along the channel dimension
and conduct a linear mapping to obtain the final output $\hat{\Xmat}_s\in\mathbb{R}^{D\times{H}\times{W}\times{C}}$ 
of spatial local window multi-head self-attention mechanism ($SLW\mbox{-}MSA$): 
\begin{equation}
  \hat{\Xmat}_s={\cal W}^R(\Wmat^P_s({\rm Concat}[\Hmat_s^1,\cdots,\Hmat_s^N])),
\end{equation}
where
$\Hmat_s^i={\rm Attention}(\Qmat_s^i,\Kmat_s^i,\Vmat_s^i)$,
$\Wmat_s^P\in\mathbb{R}^{C\times{C}}$ represents projection matrices 
and ${\cal W}^R$ represents window reverse returning to the original dimension. 
The whole process of the spatial self-attention branch can be expressed as: 
\begin{equation}
  \hat{\Xmat}_s={\rm SLW\mbox{-}MSA}(\Xmat_u).
\end{equation}

For the lack of connection between local windows, we refer to the shifted window partitioning approach of Swin Transformer \cite{ze2021video} to establish the information interaction between local windows.

\subsubsection{Temporal Self-attention Branch}
Previous reconstruction algorithms typically use 2D or 3D convolutions to explore temporal correlations.
Due to the local connection of convolutions, its receptive field is limited, 
which makes it incapable of investigating long-term correlation. 
Therefore, we exploit the long-term perception ability of Transformer to build a temporal self-attention branch. 
Different from the SSA branch, the TSA branch only performs self-attention on tokens in the same spatial position. 
In other words, the TSA branch does not calculate the spatial correlation between tokens. 

As shown in Fig.~\ref{fig:network}(d), in TSA branch, for temporal self-attention calculation, 
we reshape the input $\Xmat_{u}$ to $\Xmat_t\in\mathbb{R}^{HW\times{D}\times{C}}$. 
After this, similar to the local window self-attention calculation of the SSA branch, 
we first linearly map $\Xmat_t$ to get 
$query\;\Qmat_t\in\mathbb{R}^{HW\times{D}\times{\frac{C}{2}}}$,
$key\;\Kmat_t\in\mathbb{R}^{HW\times{D}\times{\frac{C}{2}}}$ and $value\;\Vmat_t\in\mathbb{R}^{HW\times{D}\times{\frac{C}{2}}}$: 
\begin{equation}
  \Qmat_t = \Xmat_t\Wmat_t^Q, \Kmat_t = \Xmat_t\Wmat_t^K, \Vmat_t = \Xmat_t\Wmat_t^V,
\end{equation}
where $\Wmat_t^Q\in\mathbb{R}^{C\times\frac{C}{2}},\Wmat_t^K\in\mathbb{R}^{C\times{\frac{C}{2}}}$ 
and $\Wmat_t^V\in\mathbb{R}^{C\times{\frac{C}{2}}}$ 
represent projection matrices and share parameters between different temporal windows. 

Different from the SSA branch, we decrease the channel dimensions of $\Qmat_t, \Kmat_t $ and $\Vmat_t$ 
to further reduce the computational complexity.
Then, we respectively divide $\Qmat_t$, $\Kmat_t$, $\Vmat_t$ into $N$ heads along the feature channel: 
$\Qmat_t=[\Qmat_t^1,\cdots,\Qmat_t^N]$, $\Kmat_t=[\Kmat_t^1,\cdots,\Kmat_t^N]$, $\Vmat_t=[\Vmat_t^1,\cdots,\Vmat_t^N]$, 
and the feature dimension of each head becomes $d_h=\frac{C}{2N}$. 
For each head $j=1,\cdots,N$, the attention can be calculated by the temporal window self-attention mechanism as: 
\begin{equation}
  {\rm Attention}(\Qmat_t^j,\Kmat_t^j,\Vmat_t^j)={\rm SoftMax}(\Qmat_t^{j}{\Kmat_t^{j}}^{T}/\sqrt{d}+\Bmat_t^j)\Vmat_t^j,
\end{equation}
where $\Bmat_t^j\in\mathbb{R}^{D\times{D}}$ represents the learnable relative position encoding. 
Then, we concatenate the outputs of $N$ heads along the channel dimension
and perform a linear mapping to obtain the final output $\hat{\Xmat}_t\in\mathbb{R}^{D\times{H}\times{W}\times{C}}$
of temporal window multi-head self-attention mechanism ({ $TW\mbox{-}MSA$}): 
\begin{equation}
  \hat{\Xmat}_t=\Rmat(\Wmat_t^P({\rm Concat}[\Hmat_t^1,\cdots,\Hmat_t^N]))
\end{equation}
where $\Hmat_t^j={\rm Attention}(\Qmat_t^j,\Kmat_t^j,\Vmat_t^j)$, 
$\Wmat_t^P\in\mathbb{R}^{\frac{C}{2}\times{C}}$ represents projection matrices,
and $\Rmat$ represents reshape operation. 
The whole process of the temporal self-attention branch can be expressed as: 
\begin{equation}
  \hat{\Xmat}_t={\rm TW\mbox{-}MSA}(\Xmat_u).
\end{equation}

\subsubsection{Grouping Resnet Feed Forward Network}
The FF network of a regular Transformer \cite{transformer} uses two linear mapping layers to transform features;  
the first linear layer expands the channel dimension (usually by a factor of 4), 
and the second linear layer restores the channel dimension to the original one \cite{zamir2022restormer}. 
In the whole FF network process, the operation of each feature point are independent of each other, 
and there will be no interaction between feature points. 
In this work, we modify the original FF network with details shown in the GRFF network in Fig.~\ref{fig:network}(d). 
We divide the input feature map into two parts along the channel dimension;
the first part of the feature is sent to a Resnet module, 
whose output is added to the second part of the feature, 
and the summed feature is used as input to another Resnet module. 
Following this, the outputs of the two Resnet modules are concatenated along the channel dimension 
to obtain the final output of the GRFF network. 
Given an input $\Xmat_{g}\in\mathbb{R}^{D\times{H}\times{W}\times{C}}$, 
the whole GRFF network process can be expressed as:
\begin{eqnarray}
\label{Eq:grff_b}
  \Xmat_{s1}, \Xmat_{s2} &=& {\rm Split}(\Xmat_{g}),\\ 
  \Ymat_{r1} &=& {\rm Resnet_1}(\Xmat_{s1}),\\ 
  \Ymat_{r2} &=& {\rm Resnet_2}(\Xmat_{s2}+\Ymat_{r1}),\\ 
  \hat{\Xmat}_{g} &=& {\rm Concat}([\Ymat_{r1},\Ymat_{r2}]),
  \label{Eq:grff_e}
\end{eqnarray}
where ${\rm Split}$ represents feature division along the channel, 
${\rm Concat}$ represents feature merging across the channel dimension
and $\hat{\Xmat}_{g}\in\mathbb{R}^{D\times{H}\times{W}\times{C}}$ represents the output of the GRFF network. 
Overall, the GRFF network conducts more feature mapping by grouping and using the Resnet mechanism 
to achieve multi-layer information fusion, 
and the use of convolution enhances the information interaction between adjacent feature points. 
All of these are beneficial to improve the quality of video SCI reconstruction.

\subsubsection{Whole Process of  STFormer Block}
In summary, the entire process of the STFormer block can be expressed as:
\begin{eqnarray}
  \Xmat_u &=& {\rm Norm}(\Xmat_{in}), \\
  \hat{\Xmat}_s&=& {\rm SLW\mbox{-}MSA}(\Xmat_u), \\
  \hat{\Xmat}_t&=& {\rm TW\mbox{-}MSA}(\Xmat_u), \\
  \Xmat_{st} &=& \Xmat_{in}+(\hat{\Xmat}_s + \hat{\Xmat}_t), \\
  \hat{\Xmat}_{g}&=& {\rm GRFF}({\rm Norm}(\Xmat_{st})), \\
  \Xmat_{out} &=& \Xmat_{st}+\hat{\Xmat}_{g}, 
\end{eqnarray}
where ${\rm GRFF}(\cdot)$ represents the GRFF network processing, specifically described in Eq.~\eqref{Eq:grff_b}-\eqref{Eq:grff_e}, 
${\rm Norm}(\cdot)$ represents the Layer Normalization \cite{ba2016layer}, 
and $\Xmat_{out}$ is the output of STFormer block.

\subsubsection{Computational Complexity}
In addition, we analyze the computational complexity of the spatial-temporal multi-head self-attention mechanism (${\rm ST\mbox{-}MSA}$),  
which consists of ${\rm SLW\mbox{-}MSA}$ and ${\rm TW\mbox{-}MSA}$: 
\begin{align}
  \varOmega({\rm SLW\mbox{-}MSA})&=4HWDC^2+2G_{h}G_{w}HWDC, \\
  \varOmega({\rm TW\mbox{-}MSA})&=2HWDC^2+HWD^{2}C, \\
  \varOmega({\rm ST\mbox{-}MSA})&=\varOmega({\rm SLW\mbox{-}MSA})+\varOmega({\rm TW\mbox{-}MSA}) \\
  &=6HWDC^2+2G_{h}G_{w}HWDC \notag\\
  &\quad+HWD^{2}C, 
\end{align} 
where $\varOmega(\cdot)$ represents the computational complexity, $G_h$ and $G_w$ are generally set to 7, 
and $D$ is the number of frames,  
and compare it with the global multi-head self-attention mechanism (${\rm G\mbox{-}MSA}$) \cite{cai2022mask}, 
\begin{align}
\varOmega({\rm G\mbox{-}MSA})&=4HWDC^2+2(HWD)^{2}C. 
\end{align} 
We can observe that the computational complexity of our proposed ${\rm ST\mbox{-}MSA}$ grows linearly with the spatial size $HW$, 
which is more computationally efficient than ${\rm G\mbox{-}MSA}$ (quadratic to $HW$). 

\begin{table*}[!htbp]
  \renewcommand{\arraystretch}{1.0}
  \caption{The average PSNR in  dB (left entry) and SSIM (right entry) and running time per measurement of different algorithms on 6 benchmark grayscale datasets. 
  The best results are shown in bold and the second-best results are underlined.}
  \centering
  \resizebox{\textwidth}{!}
  {
  \centering
  \begin{tabular}{c|c|c|c|c|c|c|c|c}
  \hline
  Dataset 
  & Kobe 
  & Traffic 
  & Runner 
  & Drop 
  & Crash 
  & Aerial 
  & Average 
  & Running time(s) 
  \\
  \hline
  \hline
  GAP-TV\cite{Yuan2016}
  & 26.46, 0.845
  & 20.89, 0.715
  & 28.52, 0.909
  & 34.63, 0.970 
  & 24.82, 0.838 
  & 25.05, 0.828 
  & 26.73, 0.858
  & 4.2 (CPU) \\
  \hline
  U-net\cite{Qiao2020} 
  & 27.79, 0.807
  & 24.62, 0.840
  & 34.12, 0.947
  & 36.56, 0.949
  & 26.43, 0.882
  & 27.18, 0.869
  & 29.45, 0.882
  & 0.03 (GPU)
  \\
  \hline 
  PnP-FFDNet\cite{Yuan2020c} 
  & 30.50, 0.926  
  & 24.18, 0.828
  & 32.15, 0.933
  & 40.70, 0.989
  & 25.42, 0.849
  & 25.27, 0.829
  & 29.70, 0.892
  & 3.0 (GPU)
  \\
  \hline
  PnP-FastDVDnet\cite{yuan2021plug} 
  & 32.73, 0.947
  & 27.95, 0.932 
  & 36.29, 0.962 
  & 41.82, 0.989 
  & 27.32, 0.925 
  & 27.98, 0.897
  & 32.35, 0.942
  & 6.0 (GPU)
  \\
  \hline
  DeSCI\cite{Liu2018}
  & 33.25, 0.952
  & 28.71, 0.925
  & 38.48, 0.969
  & 43.10, 0.993
  & 27.04, 0.909
  & 25.33, 0.860
  & 32.65, 0.935
  & 6180 (CPU)
  \\
  \hline
  BIRNAT\cite{Cheng2020b} 
  & 32.71, 0.950
  & 29.33, 0.942 
  & 38.70, 0.976
  & 42.28, 0.992
  & 27.84, 0.927
  & 28.99, 0.917
  & 33.31, 0.951
  & 0.16 (GPU)
  \\
  \hline
  
  GAP-net-Unet-S12 \cite{meng2020gap} 
  & 32.09, 0.944
  & 28.19, 0.929
  & 38.12, 0.975
  & 42.02, 0.992
  & 27.83, 0.931
  & 28.88, 0.914
  & 32.86, 0.947
  & 0.03 (GPU)
  \\
  \hline 
  MetaSCI \cite{Wang2021e} 
  & 30.12, 0.907
  & 26.95, 0.888
  & 37.02, 0.967
  & 40.61, 0.985
  & 27.33, 0.906
  & 28.31, 0.904
  & 31.72, 0.926
  & 0.03 (GPU)
  \\
  \hline
  RevSCI \cite{Cheng2021}
  &33.72, 0.957
  &30.02, 0.949
  &39.40, 0.977
  & 42.93, 0.992
  & 28.12, 0.937
  & 29.35, 0.924
  & 33.92, 0.956
  & 0.19 (GPU)
  \\
  \hline
  DUN-3DUnet \cite{Wu2021}
  & 35.00, 0.969
  & 31.76, 0.966
  & 40.03, 0.980
  & 44.96, 0.995
  & 29.33, 0.956
  & 30.46, 0.943
  & 35.26, 0.968
  & 1.35 (GPU)
  \\
  \hline
  \rowcolor{lightgray}
  STFormer-S
  & 33.19, 0.955
  & 29.19, 0.941
  & 39.00, 0.979
  & 42.84, 0.992
  & 29.26, 0.950
  & 30.13, 0.934
  & 33.94, 0.958
  & 0.14 (GPU)
  \\
  \hline 
  \rowcolor{lightgray}
  STFormer-B
  & \underline{35.53}, \underline{0.973}
  & \underline{32.15}, \underline{0.967}
  & \underline{42.64},\underline{ 0.988}
  & \underline{45.08},\underline{ 0.995}
  & {\bf 31.06},\underline{ 0.970}
  & \underline{31.56}, \underline{0.953}
  & 36.34, 0.974
  & 0.49 (GPU)
  \\
  \hline 
  \rowcolor{lightgray}
  STFormer-L
  & {\bf 36.02}, {\bf 0.975}
  & {\bf 32.74}, {\bf 0.971}
  & {\bf 43.40}, {\bf 0.989}
  & {\bf 45.48}, {\bf 0.995}
  & \underline{31.04}, {\bf 0.971}
  & {\bf 31.85}, {\bf 0.956}
  & {\bf 36.75}, {\bf 0.976}
  & 0.92 (GPU)
  \\
  \hline 
  \end{tabular}
  }
  \label{Tab:sim6}
\end{table*}
\begin{figure}[!h]
  \centering 
  \includegraphics[width=1.\linewidth]{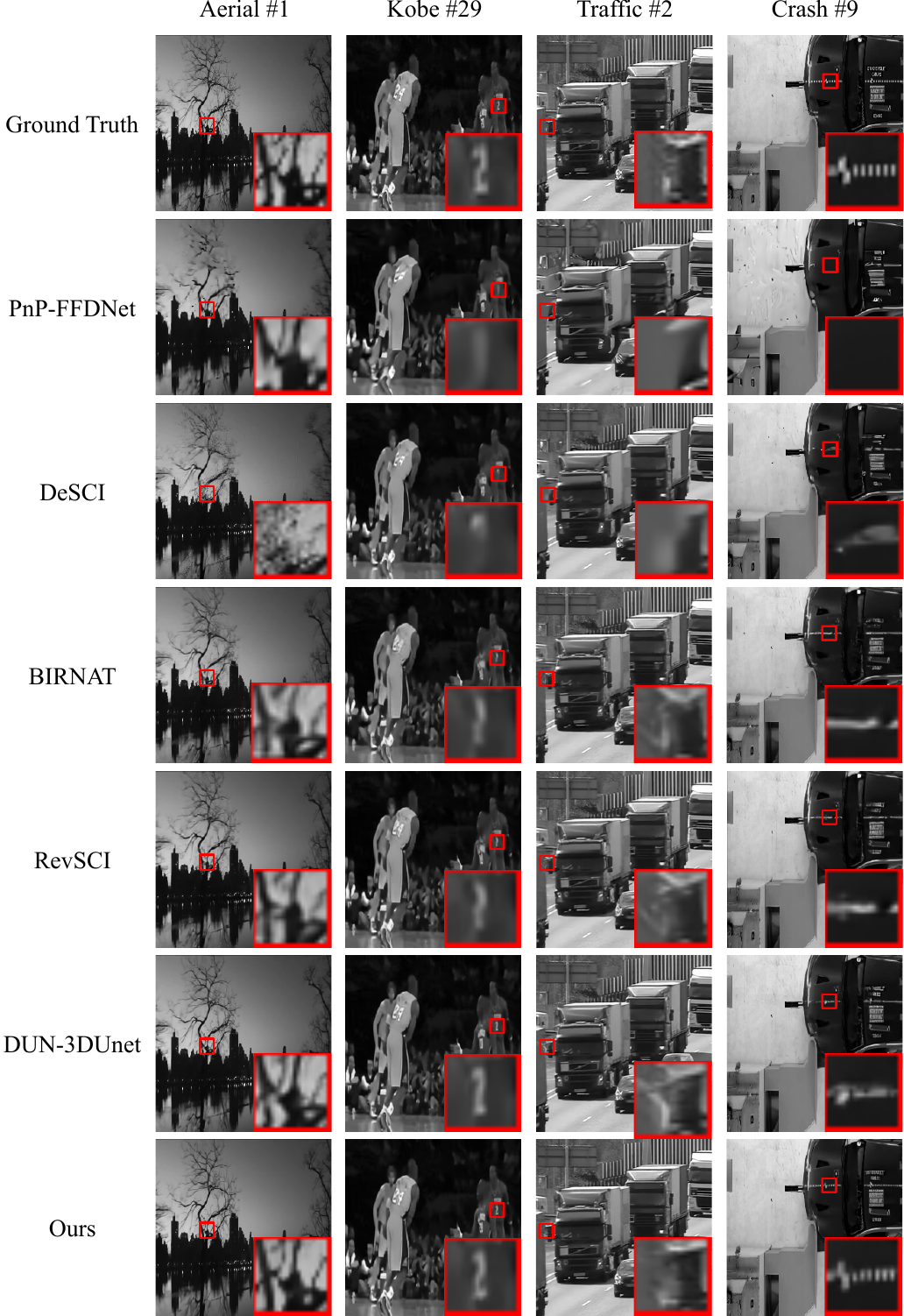}
  \caption{Comparison of reconstruction results of different reconstruction algorithms,
  PnP-FFDNet \cite{Yuan2020c}, DeSCI \cite{Liu2018}, BIRNAT \cite{Cheng2020b}, RevSCI \cite{Cheng2021}, DUN-3DUnet \cite{Wu2021}, and our proposed STFormer-B
  on several benchmark grayscale video simulation datasets (\texttt{Aerial, Kobe, Traffic, Crash}). Zoom in for better view.}
  \label{fig:sim}
\end{figure}

\section{Experimental Results \label{Sec:Result}}
In this section, we compare the performance of the proposed STFormer network 
with several SOTA video reconstruction methods on multiple simulation and real datasets. 
The peak-signal-to-noise-ratio (PSNR) and the structured similarity index
metrics (SSIM) \cite{Wang2004} are used to evaluate the performance of different video SCI reconstruction methods on simulation datasets.

\subsection{Datasets}
We use {\bf DAVIS2017} \cite{pont2017} as the training dataset for the model, 
which contains 90 different scenes with two resolutions: $480\times{894}$ and $1080\times{1920}$. 
For the grayscale simulation video testing datasets, we used six benchmark datasets 
including \texttt{Kobe, Runner, Drop, Traffic , Aerial} and \texttt{Vehicle}
with a size of $256\times{256}\times{8}$, following the setup in~\cite{Yuan2020c}. 
For the color simulation video testing datasets, we follow PnP-FastDVDnet \cite{yuan2021plug}, 
using six benchmark color simulation datasets, 
including \texttt{Beauty, Bosphorus,Jockey, Runner, ShakeNDry} and \texttt{Traffic} 
with a size of $512\times{512}\times{3}\times{8}$. 
For the large-scale simulation video testing datasets, 
we used 4 benchmark large-scale simulation datasets, including {\texttt{Messi, Hummingbird, Swinger, Football}} used in \cite{yuan2021plug}. 

For the real datasets, we used four datasets, including {\texttt{Duomino, Water Balloon, Hand}} and {\texttt{ Hammer}}, 
captured by real video SCI cameras \cite{Qiao2020,Yuan2014} for grayscale and color video, respectively. 

\subsection{Implementation Details \label{Sec:imp}}
During training, we perform data augmentation on {\bf DAVIS2017} using random horizontal flipping, random scaling, 
and random cropping. 
Following the CACTI imaging process, a series of measurements are generated. 
We use measurement and masks as inputs to train the STFormer network and 
use Adam optimizer \cite{Kingma2014} to optimize the model. 
Since our proposed STFormer network is flexible in input size, to speed up model training, 
we first train the model on data with a spatial resolution of $128\times{128}$ 
(100 epochs with an initial learning rate set be 0.0001) 
and {\em then fine-tune it} on data with a spatial resolution of $256\times{256}$ 
(20 epochs with an initial learning rate set be 0.00001). 
All experiments are run on PyTorch framework with 8 NVIDIA RTX 3090 GPUs.

\subsection{Results on Simulation Datasets}
In this section we present the results from the simulation to real dataset, first in grayscale and then in color.


To trade-off speed and performance, 
we have trained three models with different size, dubbed as STFormer-L, STFormer-B and STFormer-S, standing for Large, Base and Small networks respectively. 
The hyper-parameters of these models are as follows: 
\begin{itemize} 
  \item STFormer-S: $C$=64, block numbers = \{2,2,2,2\},
  \item STFormer-B: $C$=256, block numbers = \{2,2,2,2\},
  \item STFormer-L: $C$=256, block numbers = \{4,4,4,4\},
\end{itemize}
where $C$ represents the number of input channels of the STFormer block. 
The model parameters (Params) and theoretical computational complexity (FLOPs) 
are shown in Tab.~\ref{Tab:para_floats}, where
we can observe that the parameters and computation of our proposed STFormer-S network are less than BIRNAT and RevSCI, 
and the STFormer-B network is less than DUN-3DUnet. 

\begin{table}[!htbp]
  \renewcommand{\arraystretch}{1.0}
  \caption{Computational complexity and average reconstruction quality of 
    several SOTA algorithms on 6 grayscale benchmark datasets.}
  \centering
  {
  \centering
  \begin{tabular}{c|c|c|c|c}
  \hline
  Method & Params (M) &FLOPs (G) &PSNR &SSIM
  \\
  \hline
  BIRNAT \cite{Cheng2020b}&4.13 &390.56&33.31&0.951
  \\
  \hline
  RevSCI \cite{Cheng2021} &5.66 &766.95  &33.92 &0.956
  \\
  \hline
  DUN-3DUnet \cite{Wu2021} &61.91 &3975.83 &35.26 &0.968
  \\
  \hline
  STFormer-S &1.22 &193.47  &33.94 &0.958
  \\
  \hline
  STFormer-B &19.48 &3060.75 &36.34 &0.974
  \\
  \hline
  STFormer-L &36.81 &5363.98 &36.75 &0.976
  \\
  \hline
  \end{tabular}
  }
 \label{Tab:para_floats}
\end{table}
\subsubsection{Grayscale Simulation Video}
Currently, there are various methods for video SCI reconstruction, 
here we compare our method with some SOTA methods, 
$e.g.$, model-based iterative optimization methods (GAP-TV \cite{Yuan2016} and DeSCI \cite{Liu2018}), 
end-to-end deep learning methods (U-net \cite{Qiao2020}, MetaSCI \cite{Wang2021e}, BIRNAT \cite{Cheng2020b} and RevSCI \cite{Cheng2021}), 
plug-and-play methods (PnP-FFDNet \cite{Yuan2020c}, PnP-FastDVDnet \cite{yuan2021plug}) 
and deep unfolding methods (GAP-net \cite{meng2020gap}, DUN-3DUnet \cite{Wu2021}). 
Tab.~\ref{Tab:sim6} presents the average PSNR and SSIM values for different reconstruction methods  
on 6 benchmark grayscale datasets and the average reconstruction time for a single measurement. 
Fig.~\ref{fig:sim} shows the visualization results of several SOTA reconstruction methods. 
We can observe that our proposed method achieves higher reconstruction quality and better real-time performance
than previous SOTA methods by a large margin. 
From the visualization of reconstructed videos, our proposed method can recover more details and edge information. 
We summarize the observations in Tab.~\ref{Tab:sim6} and Fig.~\ref{fig:sim} as follows:
\begin{itemize}
    \item[1)] Our proposed method (STFormer-L) achieves an average PSNR value of 36.75 dB and an SSIM value of 0.976. 
    Compared with the previous SOTA method DUN-3DUnet (best published results) and the end-to-end deep learning method RevSCI,  
    our proposed method achieves 1.49 dB and 2.83 dB higher average PSNR, respectively.
    \item[2)] For reconstruction running time, our proposed method achieves a good balance between reconstruction quality and running performance.  
    The reconstruction quality of our proposed STFormer-B model is higher than 36 dB, and the running time is within 500ms. 
    The running speed and reconstruction quality of the STFormer-S model are higher than most current reconstruction algorithms.
    Although U-net, MetaSCI, and GAP-net run faster, 
    these algorithms have poor reconstruction quality, with an average PSNR value of less than 33 dB, which is more than 3 dB lower than STFormer-L. 
    In some complex high-speed scenarios, such as {\texttt{Traffic, Crash, Aerial}} datasets, 
    the reconstruction quality of these methods cannot even reach 29 dB. 
    \item[3)] Benefit from the powerful model capacity of Transformer 
    and its ability to effectively explore long-term dependencies,  
    STFormer has excellent performance in complex scenarios (such as {\texttt{Aerial}} data), 
    and high-speed scenarios (such as \texttt{Crash} data). 
    The reconstruction quality of these two datasets reaches 31 dB for the first time. 
    From the visualization frames in Fig.~\ref{fig:sim}, we can recover clear edges of tree trunks in \texttt{Aerial} data and the marks on the vehicles in the \texttt{Crash} data. 
    Previous SOTA methods are unable to reconstruct these details, which often leads to excessive smoothing. 
\end{itemize}
\begin{table}[!htbp]
  \renewcommand{\arraystretch}{1.0}
  \caption{Reconstruction quality of three different Transformers 
    on 6 grayscale benchmark datasets 
    showing average PSNR in dB and SSIM.}
  \centering
  \resizebox{.48\textwidth}{!}
  {
  \centering
  \begin{tabular}{c|c|c|c}
    \hline
    Dataset & TimeSformer~\cite{bertasius2021space} & Video Swin Transfomer~\cite{ze2021video} & STFormer-B
    \\
    \hline
    Kobe & 31.12, 0.932 & 30.72, 0.924 & 35.53, 0.973
    \\
    \hline
    Traffic & 27.50, 0.917 & 27.26, 0.911 & 32.15, 0.967
    \\
    \hline
    Runner & 37.13, 0.974 & 37.25, 0.971 & 42.64, 0.988
    \\
    \hline
    Drop & 40.14, 0.988 & 39.82, 0.987 & 45.08, 0.995
    \\
    \hline
    Crash &28.11, 0.931 & 28.46, 0.939 & 31.06, 0.970
    \\
    \hline
    Aerial & 28.96, 0.915 & 29.07, 0.915 & 31.56, 0.953
    \\
    \hline
    Average & 32.16, 0.943 & 32.09, 0.941 & 36.34, 0.974
    \\
    \hline

  \end{tabular}
  } 
 \label{Tab:swin_time}
\end{table}

To further explore the effectiveness of our proposed STFormer network for video SCI, we directly apply several 
SOTA video Transformer models, specifically TimeSformer \cite{bertasius2021space} and Video Swin Transformer \cite{ze2021video}, to video SCI reconstruction tasks. 
For the original Video Swin Transformer, 
since the hierarchical structure makes the spatial resolution of the model output too low, 
we use U-net \cite{Ronneberger2015} to upsample the deep features and fuse them with the shallow features. 
In this way, the model can predict the final reconstruction result. 
As shown in Tab.~\ref{Tab:swin_time}, the reconstruction quality of our proposed STFormer is significantly better than that of TimeSformer and Video Swin Transformer, 
which further verifies the effectiveness of the STFormer network for video SCI. 

\begin{table*}[!htbp]
  \renewcommand{\arraystretch}{1.0}
  \caption{The average PSNR in  dB (left entry), SSIM (right entry) and running time per measurement of different algorithms on 6 benchmark color datasets. 
  Best results are in bold and the second-best results are underlined.}
  \centering
  \resizebox{\textwidth}{!}
  {
  \centering
  \begin{tabular}{c|c|c|c|c|c|c|c|c}
  \hline
  Dataset 
  & Beauty 
  & Bosphorus 
  & Jockey 
  & Runner 
  & ShakeNDry 
  & Traffic 
  & Average 
  & Running time(s) 
  \\
  \hline
  \hline
  GAP-TV\cite{Yuan2016}
  & 33.08, 0.964
  & 29.70, 0.914
  & 29.48, 0.887  
  & 29.10, 0.878 
  & 29.59, 0.893 
  & 19.84, 0.645 
  & 28.47, 0.864 
  & 10.80 (CPU)
  \\
  \hline
  DeSCI\cite{Liu2018}
  & 34.66, 0.971 
  & 32.88, 0.952 
  & 34.14, 0.938 
  & 36.16, 0.949 
  & 30.94, 0.905 
  & 24.62, 0.839 
  & 32.23, 0.926 
  & 92640 (CPU)
  \\
  \hline
  PnP-FFDNet-gray\cite{Yuan2020c}
  & 33.21, 0.963 
  & 28.43, 0.905
  & 32.30, 0.918
  & 30.83, 0.888 
  & 27.87, 0.861
  & 21.03, 0.711
  & 28.93, 0.874
  & 13.20 (GPU)
  \\
  \hline
  PnP-FFDNet-color\cite{Yuan2020c}
  & 34.15, 0.967
  & 33.06, 0.957
  & 34.80, 0.943
  & 35.32, 0.940
  & 32.37, 0.940 
  & 24.55, 0.837 
  & 32.38, 0.931
  & 97.80 (GPU)
  \\
  \hline
  PnP-FastDVDnet-gray\cite{yuan2021plug}
  & 33.01, 0.963 
  & 30.95, 0.934 
  & 33.51, 0.928 
  & 32.82, 0.900 
  & 29.92, 0.892 
  & 22.81, 0.776 
  & 30.50, 0.899
  & 19.80 (GPU)
  \\
  \hline
  PnP-FastDVDnet-color \cite{yuan2021plug}
  & 35.27,0.972
  & 37.24, 0.978 
  & 35.63,0.950
  & 38.22, 0.965 
  & 33.71, 0.969
  & 27.49, 0.915
  & 34.60, 0.955
  & 99.05 (GPU)
  \\
  \hline 
  BIRNAT\mbox{-}color \cite{cheng2022recurrent}
  & 36.08, 0.975
  & 38.30, 0.982 
  & 36.51, 0.956 
  & 39.65, 0.973 
  & 34.26, 0.951 
  & 28.03, 0.915
  & 35.47, 0.959 
  & 0.98 (GPU)
  \\
  \hline
  \rowcolor{lightgray}
  STFormer-S
  & \underline{36.83}, \underline{0.980}
  & \underline{38.36}, \underline{0.981}
  & \underline{37.09}, \underline{0.963}
  & \underline{40.56}, \underline{0.980}
  & \underline{34.67}, \underline{0.952}
  & \underline{29.00}, \underline{0.923} 
  & \underline{36.09}, \underline{0.963}
  & 0.54 (GPU)
  \\
  \hline 
  \rowcolor{lightgray}
  STFormer-B
  & {\bf 37.37}, {\bf 0.981}
  & {\bf 40.39}, {\bf 0.988}
  & {\bf 38.32}, {\bf 0.968}
  & {\bf 42.45}, {\bf 0.985}
  & {\bf 35.15}, {\bf 0.956}
  & {\bf 30.24}, {\bf 0.939} 
  & {\bf 37.32}, {\bf 0.970}
  & 1.95 (GPU)
  \\
  \hline 
  \end{tabular}
  }
  \label{Tab:mid_color}
\end{table*}
\begin{figure*}[!h]
  \centering 
  \includegraphics[width=1.\linewidth]{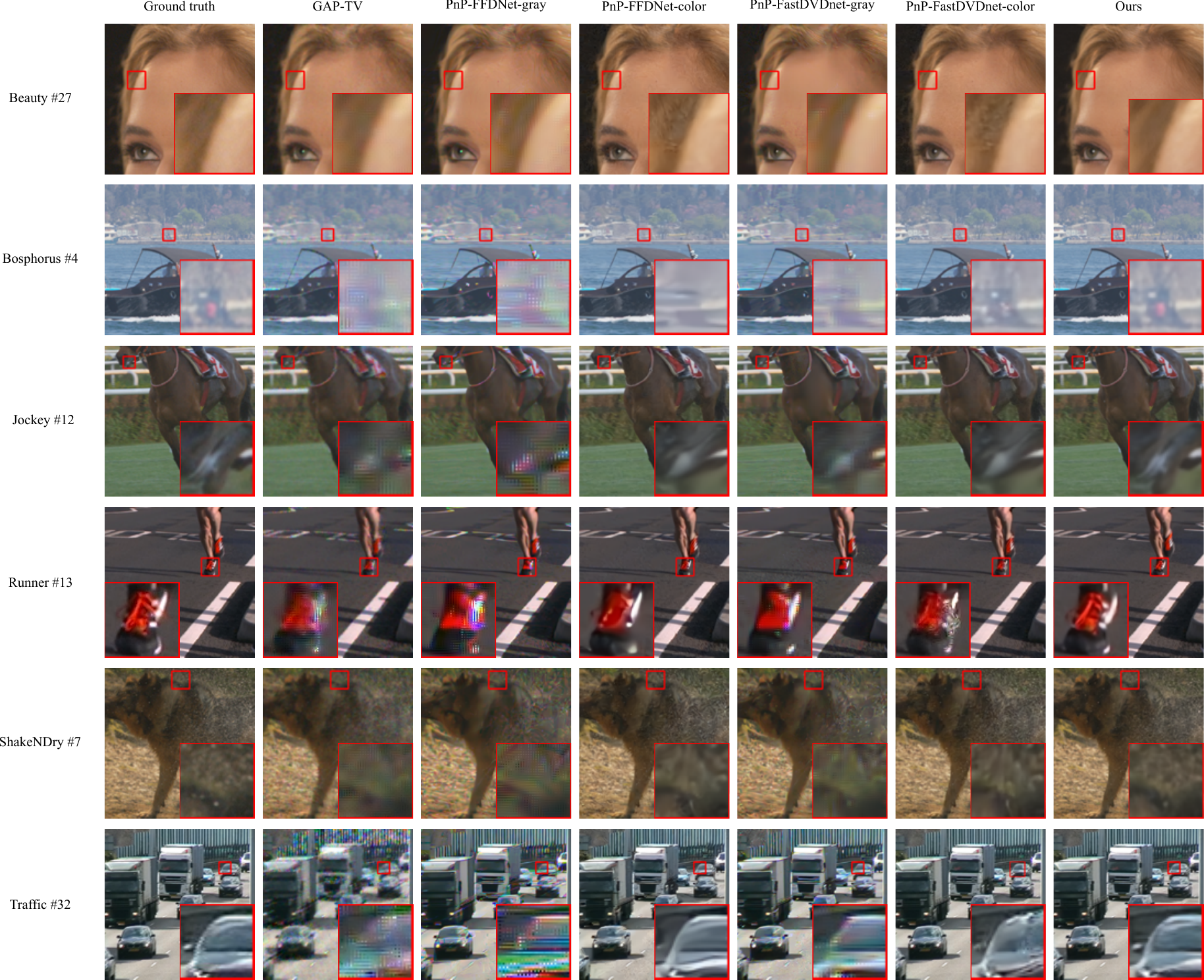}
  \caption{Comparison of reconstruction results of different reconstruction algorithms 
  (GAP-TV \cite{Yuan2016}, PnP-FFDNet-gray \cite{Yuan2020c}, PnP-FFDNet-color\cite{Yuan2020c}, PnP-FastDVD-gray \cite{yuan2021plug}, PnP-FastDVD-color\cite{yuan2021plug}, STFormer-B) 
  on several benchmark color video simulation datasets (\texttt{Beauty, Bosphorus, Jockey, Runner, ShakeNDry, Traffic}). Zoom in for a better view.}
  \label{fig:mid_color}
\end{figure*}

\subsubsection{Color Simulation Video}
To verify the effectiveness of our method on various video SCI reconstruction tasks, 
we extend the STFormer network to the color SCI reconstruction task. 
We conduct related experiments on six benchmark color RGB datasets \cite{yuan2021plug} with a spatial size of 
$512\times{512}\times{3}$, where $3$ represents the RGB channels. 
Similar to grayscale video, we compress the video with a compression rate of $B=8$. 
As shown in Fig.~\ref{fig:sci}, we capture compressed Bayer measurements using a camera with a Bayer filter. 
For each dataset with 32 color video frames, we can get 4 Bayer measurements. 

\begin{table*}[!htbp]
  \renewcommand{\arraystretch}{1.0}
  \caption{The average PSNR in dB (left entry) and SSIM (right entry) and running time per measurement of different algorithms on 4 benchmark large\mbox{-}scale datasets with a compression rate $B=8$. 
  Best results are in bold. }
  \centering
  \resizebox{\textwidth}{!}
  {
  \centering
  \begin{tabular}{c|c|c|c|c|c|c}
  \hline
  Dataset 
  & Messi 
  & Hummingbird 
  & Swinger 
  & Football 
  & Average 
  & Running time(s) 
  \\
  \hline
  \hline
  GAP-TV\cite{Yuan2016}
  & 25.20, 0.874
  & 29.64, 0.897
  & 24.64, 0.847  
  & 28.88, 0.919
  & 27.09, 0.884
  & 39.96 (CPU)
  \\
  \hline
  PnP-FFDNet\cite{Yuan2020c}
  & 30.83, 0.962 
  & 31.48, 0.945
  & 25.27, 0.881
  & 29.19, 0.930
  & 29.19, 0.930
  & 31.96 (GPU)
  \\
  \hline
  PnP-FastDVDnet\cite{yuan2021plug}
  & 31.57, 0.960
  & 33.99, 0.878
  & 26.30, 0.893
  & 34.12, 0.965
  & 31.50, 0.924
  & 209.59 (GPU)
  \\
  \hline
  \rowcolor{lightgray}
  STFormer-S
  & {\bf 33.55}, {\bf 0.964}
  & {\bf 38.20}, {\bf 0.965}
  & {\bf 31.98}, {\bf 0.964}
  & {\bf 39.13}, {\bf 0.988}
  & {\bf 35.72}, {\bf 0.970}
  & 4.25 (GPU)
  \\
  \hline 
  \end{tabular}
  }
  \label{Tab:large_color}
\end{table*}
\begin{figure*}[!htbp]
  \centering 
  \includegraphics[width=1.\linewidth]{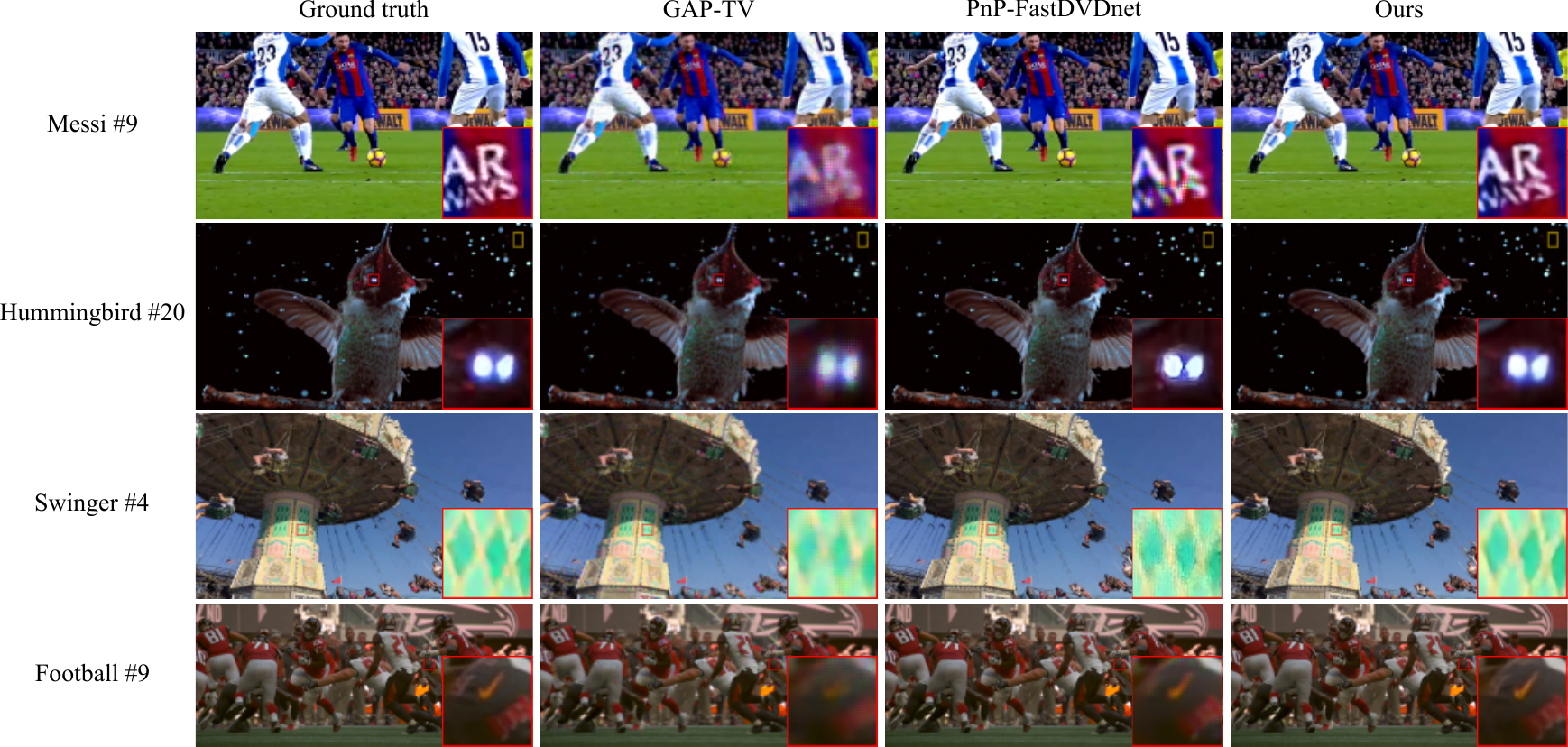}
  \caption{Comparison of reconstruction results of different reconstruction algorithms 
  (GAP-TV \cite{Yuan2016}, PnP-FFDNet \cite{Yuan2020c}, PnP-FastDVD\cite{yuan2021plug}, STFormer-S) 
  on several benchmark large-scale video simulation datasets (\texttt{Messi, Hummingbird, Swinger, Football}). Zoom in for a better view.}
  \label{fig:large_color}
\end{figure*}

Since STFormer is flexibility with respect to input size, in order to speed up training and save memory, 
we use the approach described in Sec.~\ref{Sec:imp} to train the model on small-scale data. 
Considering the inflexibility of DUN-3DUnet \cite{Wu2021} and RevSCI \cite{Cheng2021} for input size and masks, 
training a model with a spatial size of $512\times{512}$ requires  a large amount of memory or training time. 
We only compare with iterative optimization algorithms (GAP-TV \cite{Yuan2016} and DeSCI \cite{Liu2018}), end-to-end deep learning algorithm (BIRNAT\mbox{-}color \cite{cheng2022recurrent})  
and PnP algorithms (PnP-FFDNet \cite{Yuan2020c}, PnP-FastDVDnet \cite{yuan2021plug}),  
it is worth noting that PnP methods are divided into gray version and color version 
according to the use of grayscale denoiser or color denoiser. 
The reconstruction results of different algorithms are shown in Fig.~\ref{fig:mid_color} and Tab.~\ref{Tab:mid_color}, 
we can summarize the observations as follows: 
\begin{itemize}
    \item[1)] The PSNR value of the STFormer network reaches 37.32 dB, 
    which is 1.85 dB higher than the previous SOTA algorithm  BIRNAT-color, 
    especially in the high-speed motion scene {\texttt{Bosphorus}}, which is improved by 2.09 dB, 
    and it exceeds 30 dB for the first time on the {\texttt{Traffic}} dataset. 
    This shows that STFormer is also effective in color high-speed scenes. 
    \item[2)] Regarding the running time of algorithm, 
    the reconstruction of each measurement by the DeSCI algorithm is over 24 hours.  
    Although the GAP-TV and PnP algorithms achieve higher real-time performance, 
    they still take more than 10 seconds, 
    while STFormer-B reconstruction algorithm only needs 1.95 seconds, 
    which is more than 5 times faster than previous PnP  methods. 
    Recently, BIRNAT-color takes advantage of the end-to-end model to further improve the running speed of reconstruction. However, our proposed STFormer-S model achieves higher reconstruction quality and faster real-time performance. 
    \item[3)] From the visualization results, our method can recover sharper edges of 
    datasets \texttt{Beauty, Jockey,  Runner} and \texttt{Traffic}, 
    and can recover more local details of datasets \texttt{Bosphorus, ShakeNDry}. 
    The reconstructed results of GAP-TV, PnP-FFDNet-gray and PnP-FastDVD-gray methods have some artifacts, 
    while the reconstructed results of PnP-FFDNet-color and PnP-FastDVD-color methods 
    have blurred edges and serious loss of details. 
\end{itemize}

\begin{table*}[!htbp]
  \renewcommand{\arraystretch}{1.0}
  \caption{Ablation study of STFormer on 6 grayscale benchmark datasets, the average PSNR in dB and SSIM is shown.}
  \centering
  {
  \centering
  \begin{tabular}{c|c|c|c|c|c|c|c|c|c}
  \hline
  
  &TG Block
  & Video Swin Block
  & SSA Branch
  & TSA Branch
  & $MLP^2$ 
  & $MLP^1$ 
  & GRFF network
  & PSNR 
  & SSIM
  \\
  \hline
    (a) &
  & \checkmark  &  &  & \checkmark & & 
   & 33.27  & 0.952 
  \\
  \hline
  (b) &  \checkmark & \checkmark  &  &  & \checkmark & & 
   & 34.41  & 0.961
  \\
  \hline
  (c) &\checkmark  & \checkmark &  &  &  & \checkmark & 
   & 34.28 & 0.960 
  \\
  \hline
  (d) &\checkmark& & \checkmark& & \checkmark& & 
   & 33.41 & 0.954
  \\
  \hline
  (e) &\checkmark& & \checkmark& \checkmark& \checkmark& & 
   & 35.26 & 0.969 
  \\
  \hline
  (f) &\checkmark& & \checkmark & \checkmark& & \checkmark&  
  & 35.15 & 0.967
  \\
  \hline
  (g) &\ &  & \checkmark& \checkmark&  & & \checkmark
   & 35.04 & 0.964
  \\
  \hline
  (h) &\checkmark &  & \checkmark& \checkmark&  & & \checkmark
   & 36.34 & 0.974 
  \\
  \hline
  \end{tabular}
  }
 \label{Tab:ablation}
\end{table*}

\begin{table}[!htbp]
   \setlength\tabcolsep{3pt}
  \renewcommand{\arraystretch}{1.0}
  \caption{Reconstruction quality and running time (s) on 6 grayscale benchmark datasets 
    using STFormer with different number of channels and blocks}
  \centering
  {
  \centering
  \begin{tabular}{c|c|c|c|c|c}
    Model & Channel & Block & PSNR & SSIM &Running time(s)
    \\
    \hline
    STFormer-S & 64 & 8 & 33.94 & 0.958 & 0.14
    \\
    \hline
    STFormer-B & 256 & 8 & 36.34 & 0.974 & 0.49
    \\
    \hline
    STFormer-L & 256 & 16 & 36.75 & 0.976 & 0.92
    \\
    \hline

  \end{tabular}
  } 
 \label{Tab:chan_block}
\end{table}

\subsubsection{Large-scale Simulation Video}
Similar to the benchmark color data, we further extend STFormer to large-scale datasets. 
Following \cite{yuan2021plug}, we used 4 benchmark large-scale datasets, including \texttt{Messi,  Hummingbird,
Swinger} with spatial size of $1080\times{1920}\times{3}$ 
and \texttt{Football} with the spatial size of $822\times{1920}\times{3}$, 
where $3$ represents RGB channels.
Similar to the color simulation video, we generate measurements with a compression rate of $B=8$. Note that this is different from the various compression rates used in~\cite{yuan2021plug}. We set $B=8$ since we can use the same model trained for the mid-scale for the reconstruction of these large-scale datasets. This verifies the scalability and flexibility of our model. We can also train the model for large compression rates, only with additional training time and memory.

Due to the fact that network training of BIRNAT \cite{Cheng2020b} and DUN-3DUnet \cite{Wu2021} for large-scale datasets 
requires a lot of memory, while RevSCI \cite{Cheng2021} requires a longer training time, here we only compare STFormer with GAP-TV \cite{Yuan2016}, PnP-FFDNet \cite{Yuan2020c} and PnP-FastDVDnet \cite{yuan2021plug}.
Tab.~\ref{Tab:large_color} and Fig.~\ref{fig:large_color} show the reconstruction results of these algorithms. 

As shown in Fig.~\ref{fig:large_color}, we can observe that the results of GAP-TV reconstruction are blurry, 
PnP-FastDVDnet has some artifacts on some datasets, 
while our STFormer can achieve more realistic results.
More importantly, our reconstruction on these datasets can reach more than 31 dB, which proves that video SCI can be applied to real scenes.

As for the running time, since we use a small version of STFormer, 
it can provide a more efficient runtime. As shown in Tab.~\ref{Tab:large_color}, 
for the reconstruction of a single measurement, our proposed algorithm takes only 4.25 seconds,
which is 49 times faster than the previous SOTA algorithm PnP-FastDVDnet. 

\subsection{Ablation Study}
To verify the effect of each module in the proposed STFormer network on the overall reconstruction quality, 
we conducted some ablation experiments on each module. 
Tab.~\ref{Tab:ablation} shows the effect of each module on the reconstruction quality using the 6 grayscale benchmark datasets, 
where $\checkmark$ indicates that the reconstructed network includes this module, 
and no TG block indicates that the token generation method in the original Swin Transformer is used. 
In addition, $MLP^2$ and $MLP^1$ represent FF network expansion factors $\gamma=4$ and $\gamma=2$, respectively. 
We can get the following observations: 
\begin{itemize}
    \item [1)] {\bf Improvements in Token Generation Block}: 
    Tab.~\ref{Tab:ablation}(g,h) show that our proposed TG block gain is 1.30 dB higher than its counterpart. 
Similarly, combining our TG block with other Transformer blocks can also greatly 
improve the reconstruction quality (see Tab.~\ref{Tab:ablation}(a,b)). 
    \item [2)] {\bf Improvements in STFormer Block}: 
    Here, we mainly compare the STFormer block with Video Swin Transformer block. 
Tab.~\ref{Tab:ablation}(b,e) and Tab.~\ref{Tab:ablation}(c,f) show that 
our proposed STFormer block can bring an improvement of at least 0.85 dB. 
In addition, we also verified the TSA branch of the STFormer block. 
As shown in Tab.~\ref{Tab:ablation}(d,e), 
the reconstruction quality of the STFormer block with TSA branch can be improved by about 1.85 dB.

\item [3)] {\bf Improvements in Grouping Resnet Feed Forward network}:
We compare the GRFF network with the traditional FF network consisting of MLP layers. 
As shown in Tab.~\ref{Tab:ablation}(e,f,h), 
the GRFF network can provide an effective gain of more than 1.19 dB. 
   
  \item [4)] {\bf Impact of width and depth of STFormer network}:
To verify the influence of the width and depth of the STFormer network on the reconstruction quality, 
we designed three models with different number of channels used to adjust the model width 
and number of blocks used to adjust the model depth. 
As shown in Tab.~\ref{Tab:chan_block}, increasing the width and depth of the model is beneficial to 
the improvement of the reconstruction quality, 
but it also leads to a longer running time. 
\end{itemize}

\begin{table}[!htbp]
  \setlength\tabcolsep{3pt}
  \renewcommand{\arraystretch}{1.0}
  \caption{Reconstruction quality on 6 grayscale benchmark datasets 
    using random masks, the average PSNR in dB and SSIM are shown. }
  \centering
  \resizebox{.48\textwidth}{!}
  {
  \centering
  \begin{tabular}{c|c|c|c|c}
    Mask & BIRNAT  & RevSCI& DUN-3DUnet & STFormer-B
    \\
    \hline
    $Random\;Mask_1$
    & 23.15, 0.731 
    & 18.90, 0.531  
    & 31.59, 0.934 
    & 36.28, 0.974
    \\
    \hline
    $Random\; Mask_2$
    & 23.09, 0.730 
    & 18.99, 0.537 
    & 31.62, 0.934
    & 36.33, 0.974
    \\
    \hline
    $Random\;Mask_3$
    & 23.08, 0.728 
    & 18.96, 0.528 
    & 31.80, 0.935 
    & 36.32, 0.974
    \\
    \hline

  \end{tabular}
  } 
 \label{Tab:random_mask}
\end{table}
\subsection{Flexibility}
Previous end-to-end learning methods usually require retraining models 
for different masks and different spatial sizes. 
When dealing with large-scale data, they often require a large amount of memory and training time, 
which is inefficient. 
{
Transformer has a strong model capacity,
and it can dynamically adjust the attention map according to
different model inputs \cite{Liang2021a}. 
Along with our initialization method, the proposed STFormer network is robust to 
different masks. }

To verify this flexibility, 
we randomly generate three masks that are not used during training. 
As shown in Tab.~\ref{Tab:random_mask}, for different masks, 
the average PSNR value of the STFormer-B network reconstruction results remains within 0.06 dB, 
which is better than the current SOTA algorithms.
By contrast, the previous end-to-end deep learning methods BIRNAT \cite{Cheng2020b} and RevSCI \cite{Cheng2021} 
decrease more than 10 dB, 
and the deep unfolding DUN-3DUnet \cite{Wu2021} decreases more than 3 dB.

Due to this flexibility of our proposed STFormer network, 
coupled with the use of STFormer block local window and relative position bias,  
the STFormer network trained on small-scale dataset can be used for large-scale datasets. 
As shown in Tab.~\ref{Tab:mid_color} and Tab.~\ref{Tab:large_color}, 
these STFormer models are trained on small-scale data 
(spatial size less than or equal to $256\times{256}$), 
and then directly used to reconstruct data with larger spatial size, 
all achieving SOTA reconstruction results.

\begin{figure}[!htbp]
  \centering 
  \includegraphics[width=1.\linewidth]{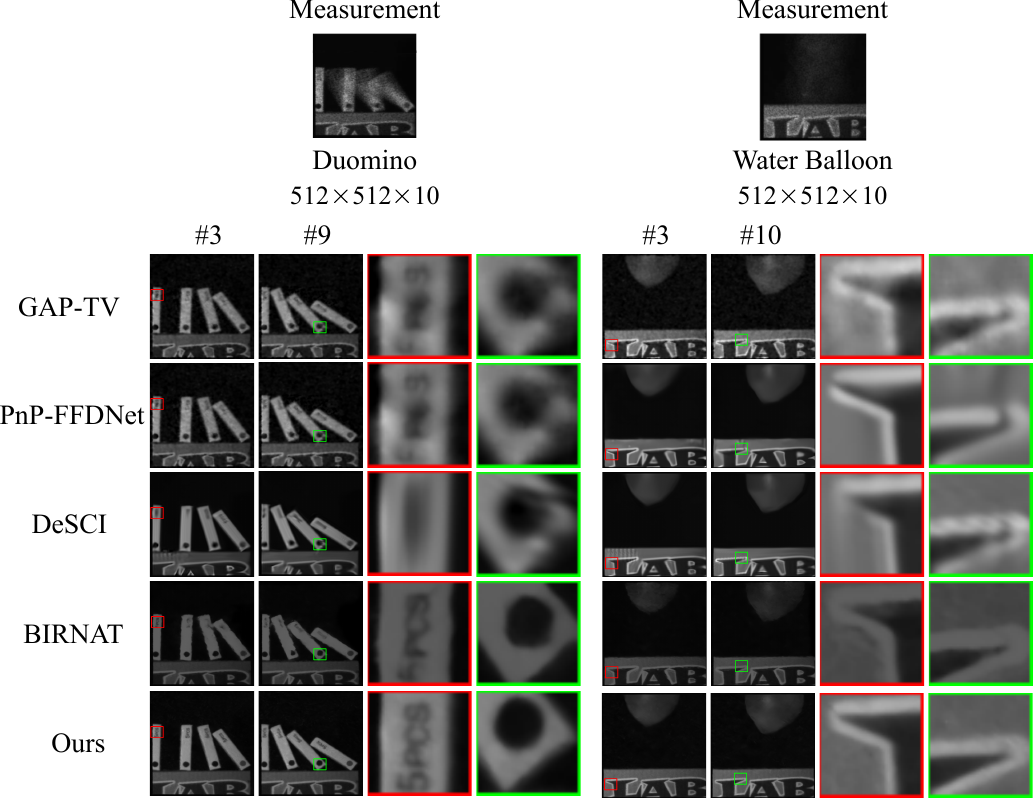}
  \caption{Comparison of reconstruction results of different reconstruction algorithms 
  (GAP-TV \cite{Yuan2016}, PnP-FFDNet \cite{Yuan2020c}, DeSCI \cite{Liu2018}, BIRNAT \cite{Cheng2020b} STFormer-B) 
  on several real datasets (\texttt{Duomino}, \texttt{Water Ballon}) with compression rate $B=10$. Zoom in for a better view.}
  \label{fig:real_cr_10}
\end{figure}

\begin{figure}[!htbp]
  \centering 
  \includegraphics[width=1.\linewidth]{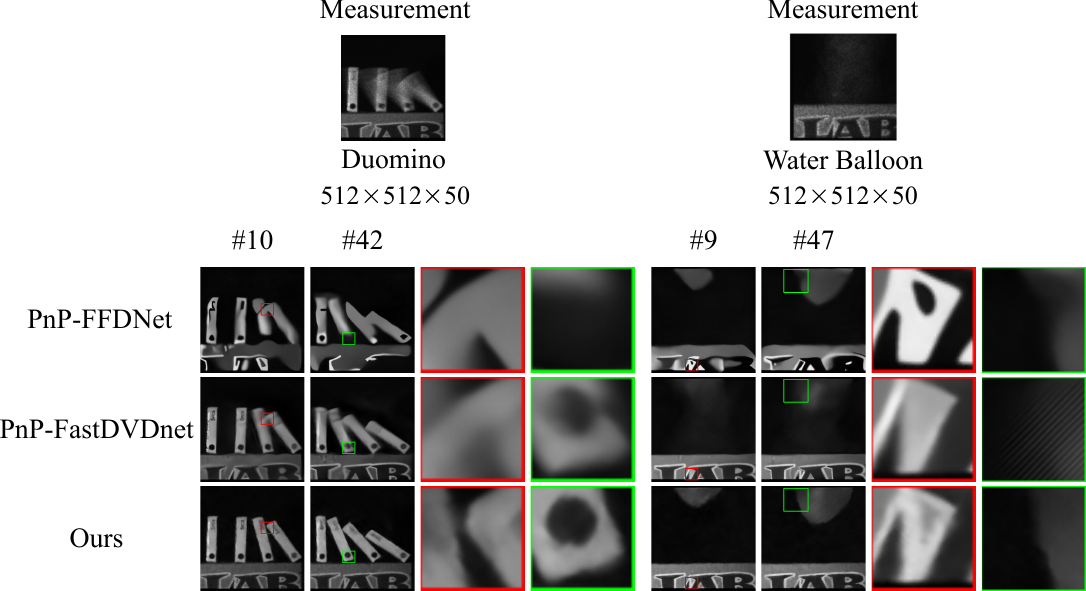}
  \caption{Comparison of reconstruction results of different reconstruction algorithms 
  (PnP-FFDNet \cite{Yuan2020c},PnP-FastDVDnet \cite {yuan2021plug} STFormer-S) 
  on several real datasets (\texttt{Duomino, Water Ballon}) with compression rate $B=50$. Zoom in for a better view.}
  \label{fig:real_cr_50}
\end{figure}

\subsection{Results on Real Video SCI Data}
We validate our proposed method on grayscale and color real data. 
Since the real video SCI imaging system has uncertain noises, 
it is more challenging to reconstruct real data.

\subsubsection{Grayscale Real Video}
For the grayscale real data, we use \texttt{Duomino, Water Ballon} and \texttt{Hand} video data captured by \cite{Qiao2020}.
It is worth noting that similar scenes are captured with different compression ratios, $i.e., B = \{10, 20, 30, 40, 50\}$
and all snapshot measurements spatial size are $512\times{512}$. 
As shown in Fig~\ref{fig:real_cr_10}, we first compared the reconstruction results with several 
SOTA reconstruction algorithms, namely GAP-TV \cite{Yuan2016}, DeSCI \cite{Liu2018}, PnP-FFDNet \cite{Yuan2020c} and BIRNAT \cite{Cheng2020b}
in the scenes of compression rate $B=10$. 
By zooming in on the local area, we can observe that our proposed algorithm can 
recover clear letters and sharp edges in the \texttt{Dumino} data and \texttt{Water Ballon} data,
while the reconstruction results of GAP-TV, PnP-FFDNet, DeSCI 
and BIRNAT algorithms over-smooth these areas with some artifacts.

\begin{figure*}[!htbp]
  \centering 
  \includegraphics[width=1.\linewidth]{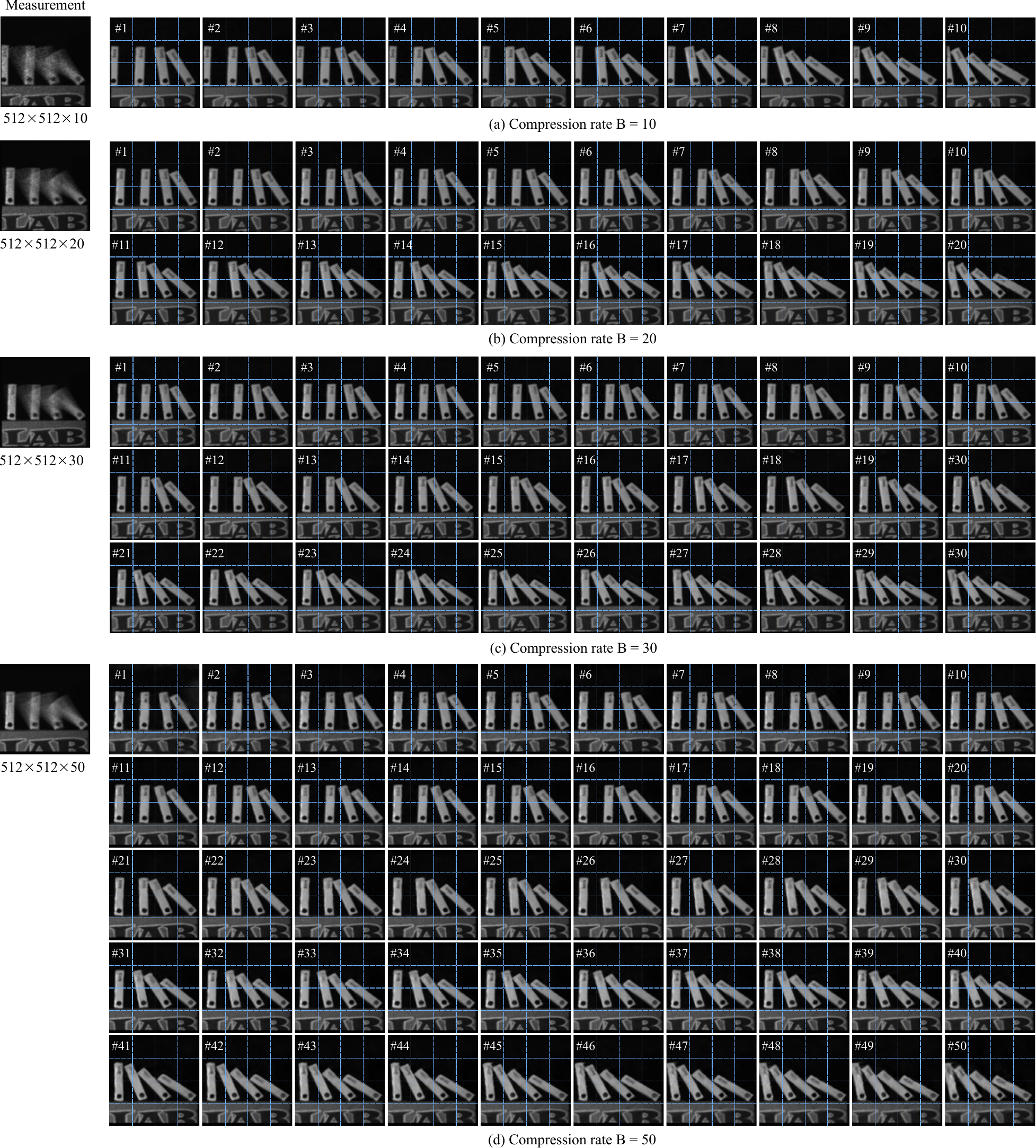}
  \caption{Reconstructed real video data (\texttt{Duomino}) with compression rate $B$ from 10 to 50, 
    Mesh is added to better visualize motion details. 
    The reconstruction algorithm is STFormer-S.}
  \label{fig:real_duomino}
\end{figure*}

\begin{figure*}[!htbp]
  \centering 
  \includegraphics[width=1.\linewidth]{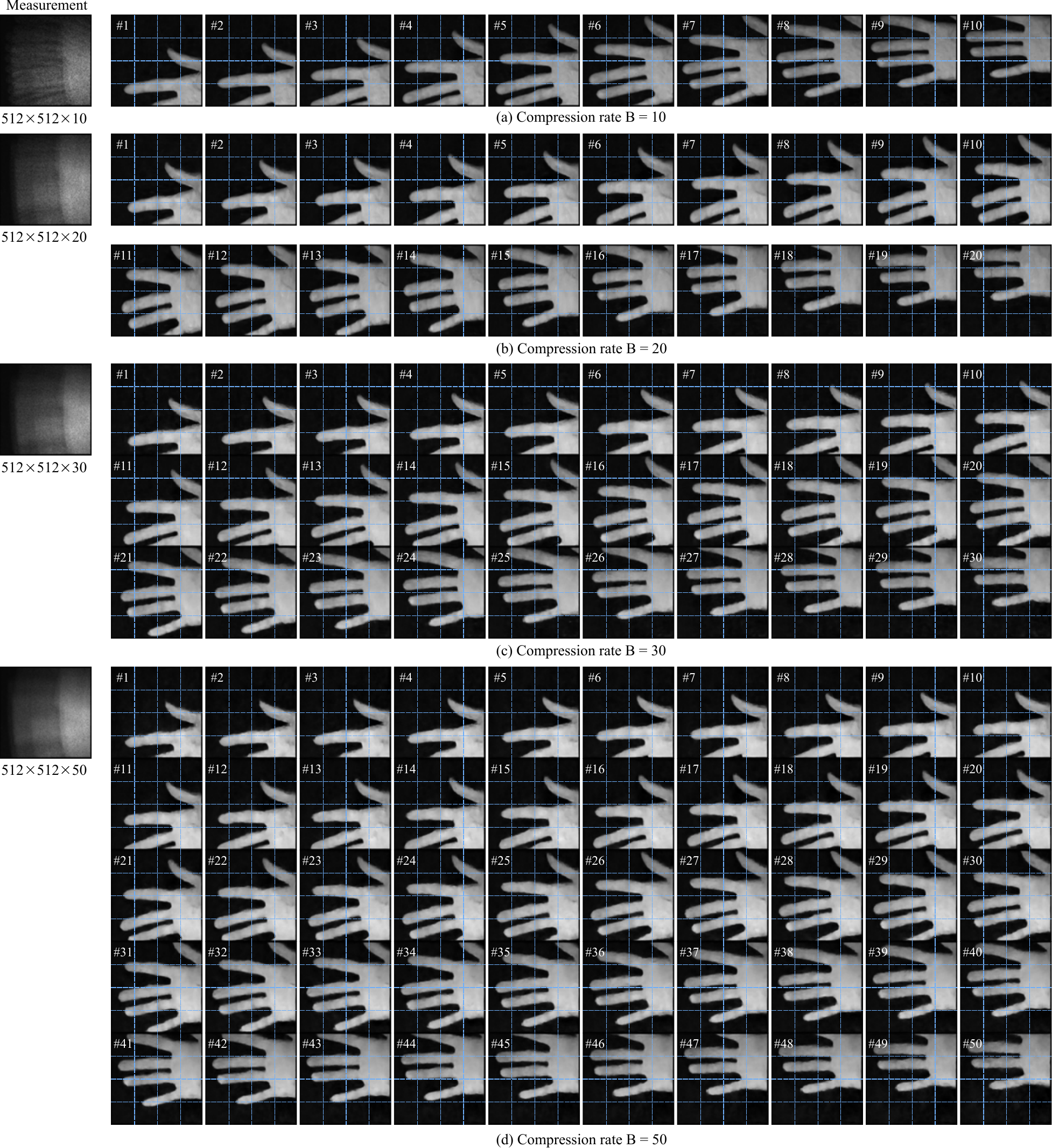}
  \caption{Reconstructed real video data (\texttt{Hand}) with compression rate $B$ from 10 to 50, 
    Mesh is added to better visualize motion details. 
    The reconstruction model is STFormer-S.}
  \label{fig:real_hand}
\end{figure*}

\begin{table}[!htbp]\tiny
  \setlength\tabcolsep{3pt}
  \caption{Running time (seconds) of real data using different algorithms }
  \centering
   \resizebox{.48\textwidth}{!}
  {
  \centering
  \begin{tabular}{c|c|c|c|c|c|c}
  \hline
  Real Data
  & Pixel resolution 
  & GAP-TV 
  & DeSCI 
  & PnP-FFDNet 
  & PnP-FastDVDnet
  & STFormer-S
  \\
  \hline
  Hand10
  & $512\times{512}\times{10}$
  & 37.8 &2880.0 &19.3 &29.5 & {\bf 1.5}
  \\
  \hline
  Hand20
  & $512\times{512}\times{20}$
  &88.7 & 4320.0 &42.4 &63.9 & {\bf 1.8}
  \\
  \hline
  Hand30
  & $512\times{512}\times{30}$
  &163.0 &6120.0 &74.7 &107.7 &{\bf 2.2}
  \\
  \hline
  Hand50
  & $512\times{512}\times{50}$
  &303.4&12600.0 &144.5 &203.9 &{\bf 2.7}
  \\
  \hline
  \end{tabular}
  }
  \label{Tab:cr50_run}
\end{table}
In addition, our proposed STFormer network can also achieve good reconstruction results 
at high compression rates, \eg, at $B=50$, which further verifies the capability of our method to explore long-term temporal dependencies. 
Although previous reconstruction algorithms can reconstruct high-compression data, 
their reconstruction results are too smooth and require a long running time (See Tab.~\ref{Tab:cr50_run}); in particular, PnP-FastDVDnet takes 3.4 minutes, DeSCI algorithm reconstruction time is more than 3 hours and and our method only needs 2.7 seconds.
Fig.~\ref{fig:real_duomino} and Fig.~\ref{fig:real_hand} show the reconstruction results 
of \texttt{Hand} and \texttt{Duomino} with $B = 10, 20, 30, 50$, respectively. 
We can observe that our proposed method can well reconstruct the desired high-speed video frames 
with compression rates from 10 to 50.
Compare with previous SOTA method PnP-FastDVDnet, our results can provide clear details of \texttt{Duomino} and the \texttt{Water Balloon} even at $B=50$. Please refer to Fig.~\ref{fig:real_cr_50}.

We have further verified our proposed algorithms on the new video SCI system built at Westlake University similar to~\cite{Qiao2020} but with different masks and different compression rates. Please refer to the reconstructed videos in the supplemental material. 

\begin{figure}[!ht]
  \centering 
  \includegraphics[width=1.\linewidth]{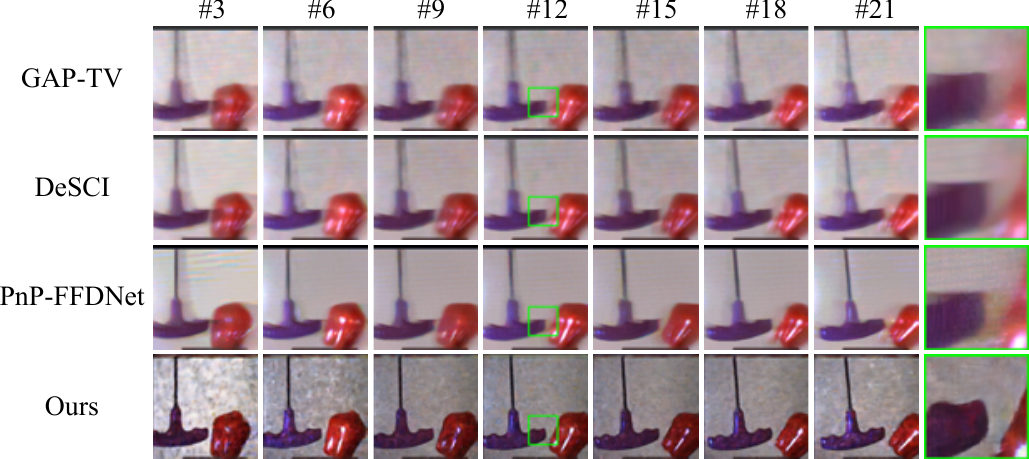}
  \caption{Comparison of reconstruction results of different reconstruction
algorithms (GAP-TV \cite{Yuan2016}, DeSCI \cite{Liu2018}, PnP-FFDNet \cite{Yuan2020c} and STFormer-S) on real color data (\texttt{Hammer}). Zoom in
for a better view.}
  \label{fig:real_color}
\end{figure}
\subsubsection{Color Real Video}
For the color real data, we use \texttt{Hammer} video data captured by \cite{Yuan2014}. 
The spatial resolution of a single Bayer mosaic measurement is $512\times{512}$ 
and the compression rate $B$ is 22. 
Since most reconstruction algorithms cannot be applied to the color data, we only compare our method with GAP-TV \cite{Yuan2016}, DeSCI \cite{Liu2018} and PnP-FFDNet \cite{Yuan2020c}.
Fig.~\ref{fig:real_color} shows the reconstruction results of these algorithms. 
By zooming in on the local areas, we can see that the reconstruction results of GAP-TV, DeSCI, and PnP-FFDNet methods have some artifacts and blurred edges, 
but our proposed STFormer method can restore these sharpe edges. 

\begin{figure*}[!htbp]
  \centering 
  \includegraphics[width=1.\linewidth]{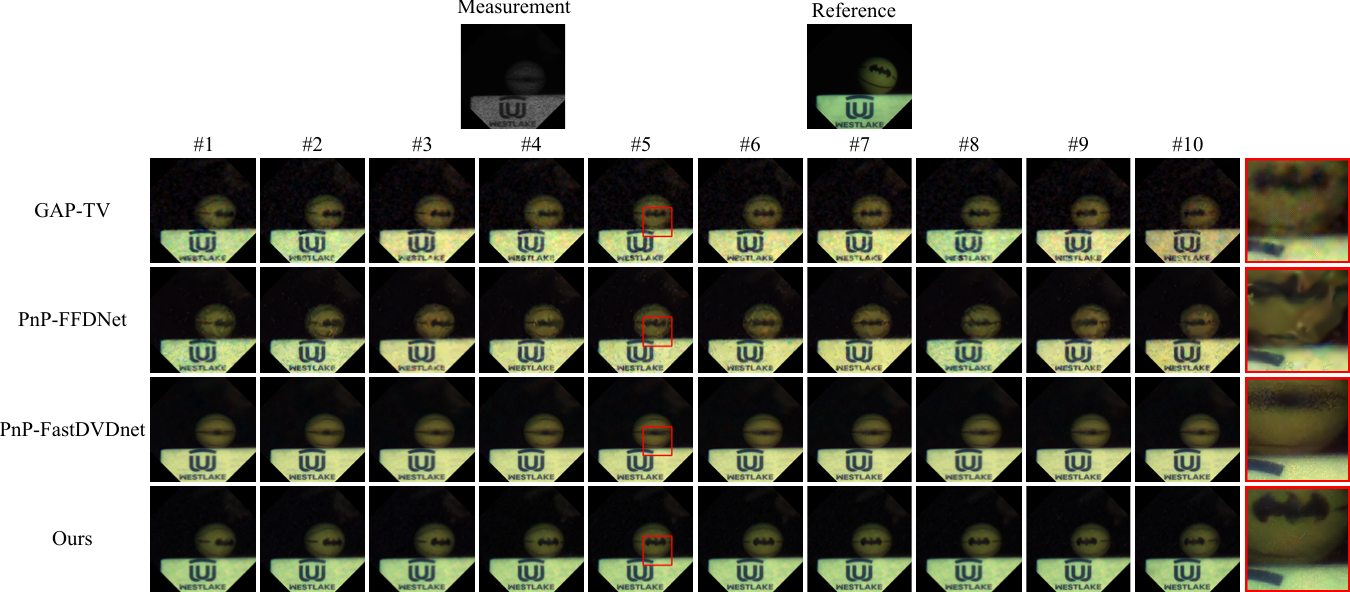}
  \caption{Comparison of reconstruction results of different reconstruction
algorithms (GAP-TV \cite{Yuan2016}, PnP-FFDNet \cite{Yuan2020c},PnP-FastDVDnet \cite{yuan2021plug} and STFormer-B) on real color data (\texttt{Ball Rotate}). Zoom in
for a better view.}
  \label{fig:westlake_ball}
\end{figure*}

\begin{figure*}[!htbp]
  \centering 
  \includegraphics[width=1.\linewidth]{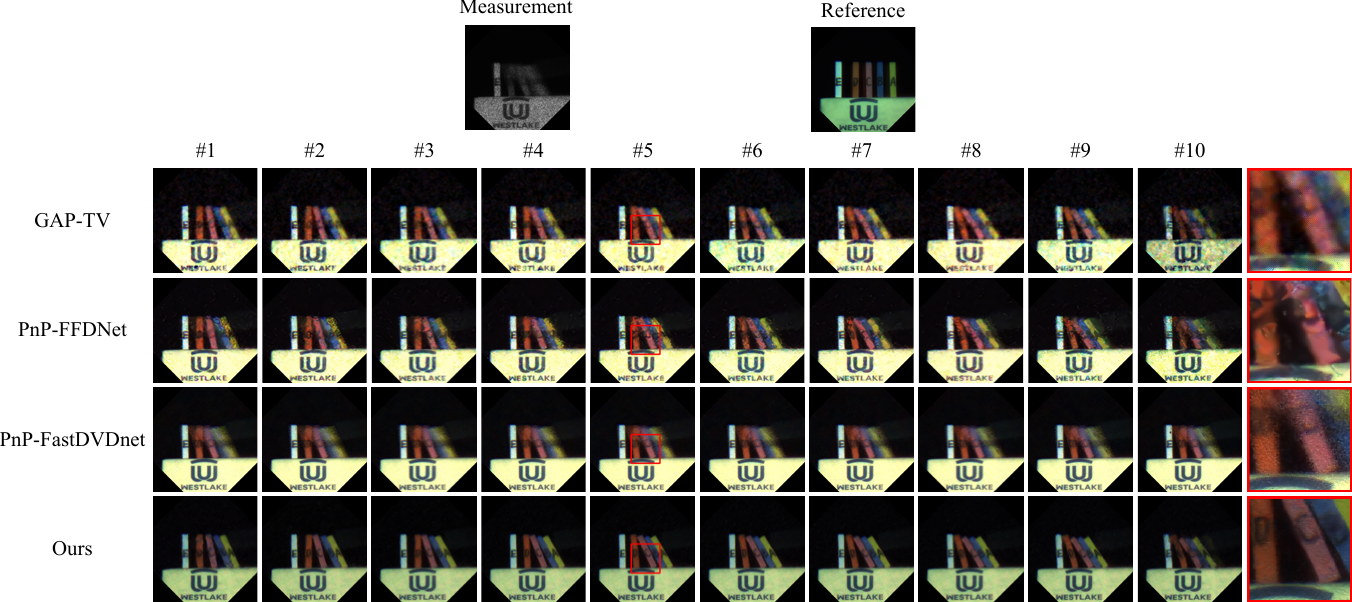}
  \caption{Comparison of reconstruction results of different reconstruction
algorithms (GAP-TV \cite{Yuan2016}, PnP-FFDNet \cite{Yuan2020c},PnP-FastDVDnet \cite{yuan2021plug} and STFormer-B) on real color data (\texttt{Duomino}). Zoom in
for a better view.}
  \label{fig:westlake_duomino}
\end{figure*}
Furthermore, to fill the gap of lacking real color data for video SCI, 
we have built a new video SCI system at Westlake University using RGB sensors 
but with different masks and different compression rates. 
The reconstructed videos are shown in Fig.~\ref{fig:westlake_ball} and Fig.~\ref{fig:westlake_duomino}. 
Comparing with previous SOTA algorithms, the reconstruction results of our proposed STFormer network are closer to the real color and can recover more details. Please refer to the enlarged areas in Figs.~\ref{fig:westlake_ball}-\ref{fig:westlake_duomino}.

\section{Conclusions  \label{Sec:Con}}
In this paper, we present STFormer, a spatial-temporal Transformer, to conduct the reconstruction task of video snapshot compressive imaging. 
Our proposed STFormer network consists of token generation block, video reconstruction block 
and a series of STFormer blocks. 
In particular, each STFormer block restricts the self-attention calculation to the spatial local window 
and time domain through the space-time factorization and local self-attention mechanism, 
which improves the computational efficiency and increases the flexibility of the model for multi-scale input.
Since STFormer can effectively explore spatial-temporal correlations, 
it achieves SOTA results on multiple video SCI reconstruction tasks. 
Especially for complex and high-speed motion scenes, 
STFormer can achieve a reconstruction quality of more than 30 dB, 
far exceeding the previous SOTA reconstruction algorithms. 
Furthermore, STFormer is the first end-to-end deep learning network with flexibility of masks and input scale, 
while enjoying fast inference, greatly facilitating applications of video SCI system in our daily life. 

Although STFormer has achieved satisfactory results on video SCI, 
the current video SCI reconstruction still faces many difficulties, 
such as the research of deep learning models suitable for different compression rates. 
In addition, for the real color data, due to the existence of complicated noise, 
there is still a huge gap between the reconstruction results and the real scene. 

Regarding future work plans, one is to extend our STFormer network to other SCI reconstruction tasks, 
such as spectral SCI  \cite{Meng2020d,wang2022snapshot,Yuan15JSTSP,Miao2019}, and the other is to use STFormer as a backbone for video action recognition \cite{wu2018compressed}, 
video object tracking \cite{zhang2021fairmot} and other tasks \cite{Lu20SEC}. 


%





\ifCLASSOPTIONcaptionsoff
  \newpage
\fi



%




\bibliographystyle{IEEEtran}
\bibliography{reference_sideinfor,reference_journal,reference_xin,reference_wangls}

\end{document}